\documentclass[12pt]{article}
\usepackage[margin=2cm]{geometry}                % See geometry.pdf to learn the layout options. There are lots.
\usepackage{graphicx}
\usepackage[sort&compress]{natbib}
\usepackage{hyperref} %Can break Arxiv submission
\AtBeginDocument{} % For correct labelling of subfiguress
\usepackage{amsmath,subfigure,enumerate,sidecap,epstopdf}
\usepackage[usenames,dvipsnames]{color}
\usepackage{amssymb,amsthm,booktabs}
\usepackage{multicol}
\usepackage{xcolor}
\usepackage{tikz}
\usepackage{float}
\usepackage{authblk}
\usepackage[font=footnotesize,labelfont=bf]{caption}

\usepackage[adobe-utopia]{mathdesign}
\usepackage[T1]{fontenc}

\setcitestyle{authoryear}
\makeatletter
\renewcommand{\section}{\@startsection
{section}%                   % the name
{1}%                         % the level
{0mm}%                       % the indent
{-\baselineskip}%            % the before skip
{0.5\baselineskip}%          % the after skip
{\normalfont\Large\bf}} % the style
\renewcommand{\subsection}{\@startsection
{subsection}%                   % the name
{2}%                         % the level
{0mm}%                       % the indent
{-\baselineskip}%            % the before skip
{0.5\baselineskip}%          % the after skip
{\normalfont\large\bf}} % the style
\renewcommand{\subsubsection}{\@startsection
{subsubsection}%                   % the name
{3}%                         % the level
{0mm}%                       % the indent
{-\baselineskip}%            % the before skip
{-\baselineskip}%          % the after skip
{\normalfont\bf}} % the style
\makeatother

\newcommand{\D}{\mathrm{d}}
\renewcommand{\vec}[1]{\boldsymbol{#1}}

\newmuskip\pFRqskip
\mathchardef\pFRcomma=\mathcode`, % keep a copy of the comma

\newcommand*\pFRq[5]{%
  \begingroup
  \begingroup\lccode`~=`,
    \lowercase{\endgroup\def~}{\pFRcomma\mkern\pFRqskip}%
  \mathcode`,=\string"8000
  {}_{#1}\boldsymbol{\mathrm{F}}_{#2}\biggl[\genfrac..{0pt}{}{#3}{#4};#5\biggr]%
  \endgroup
}

\newmuskip\pFqskip
\mathchardef\pFcomma=\mathcode`, % keep a copy of the comma

\newcommand*\pFq[5]{%
  \begingroup
  \begingroup\lccode`~=`,
    \lowercase{\endgroup\def~}{\pFcomma\mkern\pFqskip}%
  \mathcode`,=\string"8000
  {}_{#1}\mathrm{F}_{#2}\biggl[\genfrac..{0pt}{}{#3}{#4};#5\biggr]%
  \endgroup
}
\title{Localisation for a line defect in an infinite square lattice}

\author[1]{D.J. Colquitt}
\author[2]{M.J. Nieves}
\author[2]{I.S. Jones}
\author[1]{A.B. Movchan}
\author[1]{N.V. Movchan}
\affil[1]{Department of Mathematical Sciences, University of Liverpool, Liverpool L69 3BX, U.K.}
\affil[2]{School of Engineering, John Moores University, James Parsons Building, Byrom Street, Liverpool L3 3AF, U.K.}
\date{}

\begin{document}

\maketitle

\begin{center}
\footnotesize{\textit{
Published version appears in Proceedings of the Royal Society of London Series A\\
\url{http://rspa.royalsocietypublishing.org/} \\
Proc R Soc A 469: 20120579\\
(\url{http://dx.doi.org/10.1098/rspa.2012.0579})
}}
\end{center}

\abstract{
Localised defect modes generated by a finite line defect composed of several masses, embedded in an infinite square cell lattice, are analysed using the linear superposition of Green's function for a single mass defect.
Several representations of the lattice Green's function are presented and discussed.
The problem is reduced to an eigenvalue system and the properties of the corresponding matrix are examined in detail to yield information regarding the number of symmetric and skew-symmetric modes.
Asymptotic expansions in the far field, associated with long wavelength homogenisation, are presented.
Asymptotic expressions for the Green's function in the vicinity of the band edge are also discussed.
Several examples are presented where eigenfrequencies linked to this system and the corresponding eigenmodes are computed for various defects and compared with the asymptotic expansions.
The case of an infinite defect is also considered and an explicit dispersion relation is obtained.
For the case when the number of masses within the line defect is large, it is shown that the range of the eigenfrequencies can be predicted using the dispersion diagram for the infinite chain. 
}

\section{Introduction}

Despite being first studied in the late 17\textsuperscript{th} century by~\cite{Newton}, wave propagation through discrete structures remains an active area of research today.
A well-known and interesting feature of discrete media is the existence of pass and stop bands.
The present paper examines the effect of a finite line of defects in an infinite square lattice.
The behaviour of a lattice with a single point defect, or point source, can be described by the lattice Green's function.
Such Green's functions have been studied by~\cite{martin2006} for the two-dimensional square lattice.
The resulting solution  was analysed for frequencies within the pass band and the corresponding asymptotics at infinity were also obtained. 

\cite{movchan-slepyan} examined several classes of continuous and discrete models with various forcing or defect configurations.
When the forcing frequency, or natural frequency of the defect, is  located in the stop band, localised  modes were identified.
For a particular choice of the mass variation, these defect modes  can then be linked to stop-band Green's kernel which can be used in the construction of the defect modes as discussed.

Using similar methods, \cite{Gei2009} considered the effect of uniform pre-stress on the propagation of flexural waves through an elastic beam on a Winkler foundation.
Particular attention was devoted to band-gap localised modes and control of the position of stop-bands via pre-stress.
It was found that a tensile pre-stress can increase the frequency at which a particular band-gap occurs.
Alternatively, band-gaps can be \emph{annihilated} with the application of an appropriate pre-stress.

Lattice Green's functions are often studied in isolation and have proved a rich area of research (see, for example,~\cite{Joyce2001},~\cite{Delves2007},~\cite{Zucker2011}, and references therein).
For $d$-dimensional lattices, the Green's function is typically expressed as a $d$-dimensional Fourier integral.
It is often possible to evaluate one or more of the integrals, as in the paper by~\cite{movchan-slepyan}, but for $d>1$ the Green's function cannot be expressed in terms of elementary functions.
In the present article, several different representations of the square lattice Green's function are presented, which prove useful for band edge expansions.

Classical applications in the theory of defects in crystals and dislocations follow from the  fundamental work of~\cite{maradudin1965}, where explicit closed form solutions were derived for a heterogeneous lattice system when two distant particles of different masses are interchanged. 
More recently, the envelope function based perturbation approach was developed by~\cite{mahmoodian2009} and~\cite{dossou2008} for analysis of waveguides in photonic crystal structures.
In the latter case, an array of cylinders (inclusions) represents a waveguide within a two-dimensional structure, and the frequencies of the guided modes are close to the band edge of the unperturbed doubly periodic system. 

Localisation of waves due to an infinite line defect embedded in an infinite square lattice, has been considered by~\cite{Osharovich_etal}. 
For the case of an infinite line defect, dispersion relations can be computed in explicit form allowing spatially localised waveguide modes to be analysed.

\cite{Slepyan2002} presents a detailed discussion of applications for dynamic lattice problems involving cracks modelled as semi-infinite faults, for both square and triangular elastic lattices.
For a structured interface and a crack  propagating with constant speed within a square lattice, localised modes were analysed by~ \cite{Mishuris_etal}.
In particular, it was shown that the crack propagation can be supported  by a  sinusoidal wave localised along the crack, which the authors refer to as a \emph{knife wave}.
Using the lattice model, the dispersion relations for the crack within the square lattice can be derived.
As shown in numerical experiments, these relations  allow for the prediction of the average crack speed within the lattice when a fracture criterion for the crack path bonds is introduced.
More recently, \cite{Nieves2012} studied the propagation of a semi-infinite dynamic crack in a non-uniform elastic lattice. The crack stability was analysed and it was shown that information regarding unstable crack growth could be obtained from the study of the steady state regime.
 
For the finite-frequency regime, a theory of asymptotic homogenisation for scalar lattices has been implemented by \cite{Craster2010}.
This theory makes use of information related to standing wave modes found in the lattice problem.
Then a two-scale asymptotic procedure can be applied  in order to obtain an effective partial differential equation for the corresponding macroscale that contains information about the microscale structure.
 
\cite{Ayzenbergstepanenko2008} showed that point forces acting at \emph{saddle-point} frequencies within square and triangular lattices produce localised primitive wave forms within the lattice.
Similar localised primitive waveforms were demonstrated for the in-plane motion of elastic lattices by~\cite{Colquitt2011}.
The shape of these waveforms, created by a point force, were linked to the dispersive properties of Bloch waves in the lattice.

The structure of this paper is as follows.
In section \ref{sec:finite}, the problem of a finite line of defects (created by a perturbation of point masses) embedded in an infinite square  lattice is considered.
Several representations for the Green's matrix are presented, including integral forms and representation in terms of a generalised hypergeometric function.
Localised defect modes for the finite line are analysed in section~\ref{sec:finite}(\ref{sec:loc-mod}).
Therein, the necessary and sufficient condition for the existence of localised modes is formulated, and asymptotic expansions in the far field are also presented.
Band edge expansions are constructed using an analytic continuation of the Green's function.
Illustrative examples for a finite number of defects are given in section \ref{sec:illustex}, where eigenfrequencies and eigenmodes are presented and compared with asymptotic results from the previous section.
For the finite line defect it is observed that, in contrast to the 1D and 3D cases, a localised defect mode may be initiated by removing any amount of mass from a line of nodes in the lattice.
The analysis of a finite-sized defect region is accompanied by the waveguide modes that may exist in a lattice containing an infinite chain of  point masses, as in section \ref{sec:infinite}.
The governing equations for such a waveguide and the solvability of the problem are discussed in section \ref{sec:infinite}.
In section \ref{sec:infinite}(\ref{sec:sym}), the dispersion relation corresponding to the localised mode for the infinite chain is given.
Finally, in section~\ref{sec:numsim}, a numerical simulation illustrates  that the solution for the problem of the infinite chain can be used to predict the range of eigenfrequencies of localised modes for a finite but sufficiently long  array of masses representing a rectilinear defect in a square lattice.

\section{A finite inclusion in an infinite square lattice}
\label{sec:finite}
\begin{SCfigure}[1][htb]
\centering
\includegraphics[scale=0.8]{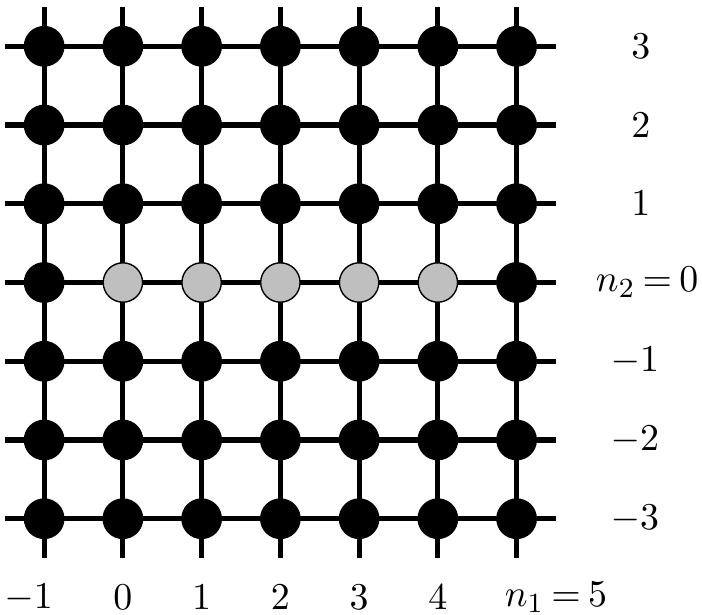}
\caption{\label{fig:lattice}A finite line of defects in an infinite square lattice.
The length of the links, the stiffness of the bonds and the mass of the black nodes are taken as natural units.}
\end{SCfigure}

Consider a square meshing of $\mathbb{R}^2$ such that each node is labelled by the double index $\vec{n}\in\mathbb{Z}^2$.
Let there be $N$ defects (with $N\in\mathbb{N}$) distributed along $n_2=0$ as shown in figure~\ref{fig:lattice}.
The defects are characterised by a non-dimensional mass $0< r < 1$, where the mass of the ambient nodes is taken as a natural unit.
The stiffness and lengths of the lattice bonds are uniform and taken as further natural units.
All physical quantities, such as the frequency and displacement, have been normalized according to these natural units and are therefore dimensionless.
Let $u_{\vec{n}}$ denote the complex amplitude of the time-harmonic out-of-plane displacement of node $\vec{n}$.
Then, the equation of motion is
\begin{equation}
\label{eq:gov}
\begin{split}
u_{\vec{n}+\vec{e}_1} + u_{\vec{n}-\vec{e}_1} + u_{\vec{n}+\vec{e}_2} + u_{\vec{n}-\vec{e}_2} + (\omega^2 - 4)&u_{\vec{n}}\\
& =
(1-r)\omega^2\delta_{0,n_{2}}\sum_{p=0}^{N-1}u_{\vec{n}}\delta_{p,n_{1}},
\end{split}
\end{equation}
where $\omega$ is the radian frequency, $\vec{e}_i = [\delta_{1,i},\delta_{2,i}]^\mathrm{T}$, and $\delta_{i,j}$ is the Kronecker Delta. By means of the discrete Fourier Transform
\begin{equation}
\mathcal{F}: u_{\vec{n}} \mapsto u^\mathrm{FF}(\vec{\xi}) = \sum_{\vec{n}\in\mathbb{Z}^2} u_{\vec{n}}e^{-i\vec{n}\cdot\vec{\xi}},
\end{equation}
the governing equation~\eqref{eq:gov} may be written
\begin{equation}
(\omega^2 - 4 + 2\cos \xi_1 + 2\cos \xi_2)u^\mathrm{FF}(\vec{\xi}) =  (1-r)\omega^{2}
\sum_{p=0}^{N-1}u_{p,0}e^{-i p\xi _1}.
\label{eq:trans-field}
\end{equation}
The positive root of the parenthesised term represents the dispersion equation for the ambient lattice.
It is observed that for $\omega^2>8$ there exist no real solutions to the dispersion equation.
Hence, the ambient lattice possesses a semi-infinite stop band: $\omega^2\in(8,\infty)$.
Inverting the transform yields the discrete field
\begin{equation}
u_{\vec{n}}(\omega) = (1-r)\omega^2\sum_{p=0}^{N-1} u_{p,0}g(\vec{n},p;\omega),
\label{eq:field}
\end{equation}
where $g(\vec{n},p;\omega)$ is the shifted Green's matrix defined as:
\begin{equation}
g(\vec{n},p;\omega) = \frac{1}{\pi^2}\int\limits_0^\pi\int\limits_0^\pi \frac{\cos\left([n_1-p]\xi_1\right)\cos(n_2\xi_2)}{\omega^2-4+2\cos\xi_1+2\cos\xi_2}\D\xi_1\D\xi_2.
\label{eq:2D-Green}
\end{equation}
For the purposes of numerical evaluation and asymptotic analysis in the stop band of the ambient lattice ($\omega^2>8$), it is convenient to rewrite the Green's matrix as a single integral
\begin{equation}
g(\vec{n},p;\omega) = \frac{1}{2\pi}\int\limits_0^\pi\frac{(\sqrt{a^2-1}-a)^{|n_{1}-p|}}{\sqrt{a^2-1}}\cos\left(n_{2}\xi_2\right) \D\xi_2,
\label{eq:sgfx1}
\end{equation}
where $a=\omega^2/2-2+\cos\xi_2$.
Reversing the order of integration yields the same result, but with $n_{1}-p$ and $n_{2}$ interchanged, and $\xi_{1}$ interchanged with $\xi_{2}$.
An alternative representation can be found in the book by~\cite{van-der-pol}  as
\begin{equation}
g(\vec{n},p;\omega) = \frac{(-1)^{n_{1}-p+n_{2}}}{2}\int\limits_0^\infty I_{n_{1}-p}(x)I_{n_{2}}(x)e^{-\alpha x}\D x,
\label{eq:g-bessel}
\end{equation}
where $I_m(x)$ is the modified Bessel function of the first kind, $\alpha = \omega^2/2-2>2$.
The integral is symmetric about $n_1-p=0$ and $n_2=0$ and therefore it may be assumed, without loss of generality, that $n_1\geq p$ and $n_2\geq0$.
The integral~\eqref{eq:g-bessel} may then be represented in terms of a regularised generalised hypergeometric function (see~\cite{prudnikov-v4}, section 3.15.6, equation 8)
\begin{equation}
g(\vec{n},p;\omega) = \frac{(-1)^{m+n_2}}{(2\alpha)^{1+m+n_2}}((m+n_2)!)^2\pFRq{4}{3}{a_1, a_1,a_2,a_2}{b_1,b_2,b_1+b_2-1}{\frac{4}{\alpha^2}},
\label{eq:g-hypergeom}
\end{equation}
where $m=n_1-p$, $a_1=(1+m+n_2)/2$, $a_2=(2+m+n_2)/2$, $b_1=1+m$, and $b_2=1+n_2$.
The series~\eqref{eq:g-hypergeom} is convergent for $\alpha^2>4$, that is, everywhere in the stop band of the ambient lattice.
It is observed that along the ray $m=n_2$, the Green's matrix may be written in terms of Gauss' hypergeometric function.
In particular, equation~\eqref{eq:g-hypergeom} reduces to
\begin{equation}
g(n,n,0;\omega) = \frac{((2n)!)^2}{(2\alpha)^{1+2n}}\pFRq{2}{1}{1/2+n, 1/2+n}{1+2n}{\frac{4}{\alpha^2}}.
\label{eq:g-hypergeom-mm}
\end{equation}
The function~\eqref{eq:g-hypergeom-mm} is strictly positive in the region $n\geq0$ and $\alpha>2$.
Hence, for a single defect, the lattice nodes along the diagonal rays do not oscillate relative to each other.

Furthermore, for the case of $m=n_2=0$, the integral representation~\eqref{eq:2D-Green} reduces to the $2$-fold Watson integral (see, for example, \cite{Joyce2001} and \cite{Zucker2011}).
Using a simple change of variables~\eqref{eq:2D-Green} may be written in terms of an elliptic integral, or alternatively, one may use~\eqref{eq:g-hypergeom-mm} and observe that
\begin{equation}
g(0,0,0;\omega) = \frac{1}{2\alpha}\pFRq{2}{1}{1/2, 1/2}{1}{\frac{4}{\alpha^2}} =
\frac{1}{\alpha\pi}K\left(\frac{4}{\alpha^2}\right),
\label{eq:g00}
\end{equation}
where $K(x)$ is the complete elliptical integral of the first kind.
Together with equation~\eqref{eq:g00}, the representation~\eqref{eq:g-bessel} is particularly useful since, by repeated integration by parts and use of the identity $I_{n}(x)=2I^{\prime}_{n-1}(x) - I_{n-2}(x)$, one can iterate from $g(0,0,0;\omega)$ to a general $g(\vec{n},p;\omega)$.

\subsection{Localised modes}
\label{sec:loc-mod}

Of primary interest are localised modes, that is, modes of vibration at frequencies that are not supported in the ambient lattice and therefore decay rapidly away from the defect sites.
Introducing the vector $\mathcal{U}=[u_{0,0},u_{2,0},\ldots,u_{N-1,0}]^\mathrm{T}$ and choosing $n_2=0$ in equation~\eqref{eq:field} yields the eigenvalue problem
\begin{equation}
\mathcal{U} = (1-r)\omega^2\mathcal{G}(\omega)\mathcal{U},
\label{eq:spectral}
\end{equation}
where the matrix entries $[\mathcal{G}(\omega)]_{ij} = g(i-1,0, j-1;\omega)$.
Clearly, $\mathcal{G}$ is symmetric and Toeplitz (and hence bisymmetric and centrosymmetric)
\begin{equation}
\mathcal{G} = \begin{pmatrix}
\mathcal{G}_{11} & \mathcal{G}_{12} & \mathcal{G}_{13} & \cdots & \mathcal{G}_{1(N-1)} & \mathcal{G}_{1N} \\
	 & \mathcal{G}_{11} & \mathcal{G}_{12} & \cdots & \mathcal{G}_{1(N-2)} & \mathcal{G}_{1(N-1)} \\
	& 	& \mathcal{G}_{11} & \cdots & \mathcal{G}_{1(N-3)} & \mathcal{G}_{1(N-2)} \\
	&		&	& \ddots & \vdots & \vdots\\
	&		&	&		& \mathcal{G}_{11} & \mathcal{G}_{12} \\
	&		&	&		&				& \mathcal{G}_{11}
\end{pmatrix},
\end{equation}
which greatly reduces the number of required computations.
Indeed, for $N$ defects the matrix $\mathcal{G}$ has $N$ independent elements.
The solvability condition of the spectral problem~\eqref{eq:spectral} yields a transcendental equation in $\omega$,
\begin{equation}
\det\left[\mathbb{I}_{N}-(1-r)\omega^2\mathcal{G}\right] = 0,
\label{eq:trans}
\end{equation}
where $\mathbb{I}_{N}$ is the $N\times N$ identity matrix.
Equation~\eqref{eq:trans} is the necessary and sufficient condition for the existence of a localised mode.
Symmetry implies that there exists an orthonormal set of $N$ eigenvectors of $\mathcal{G}$ and hence, $N$ eigenvalues (frequencies).
The centrosymmetry of $\mathcal{G}$ allows the number of symmetric and skew-symmetric modes to be determined (see, for example,~\cite{cantoni}).
Introducing the $N\times N$ exchange matrix
\begin{equation}
\mathbb{J}_{N} = \begin{pmatrix}
0 & 0 & 0 & 1\\
0 & 0 & 1 & 0\\
0 & \iddots & 0 & 0\\
1 & 0 & 0 & 0
\end{pmatrix},
\end{equation}
an eigenmode is said to be symmetric if $\mathcal{U} = \mathbb{J}_{N}\mathcal{U}$ and skew-symmetric if $\mathcal{U} = -\mathbb{J}_{N}\mathcal{U}$. For a system of $N$ defects there exist $\lceil N/2 \rceil$ symmetric modes and $\lfloor N/2 \rfloor$ skew-symmetric modes, where $\lceil\cdot\rceil$ and $\lfloor\cdot\rfloor$ are the ceiling and floor operators respectively.
Of course here, symmetry refers to the symmetry of the eigenmodes in the $n_{1}$ direction about the centre of the defect line. Due to the symmetry of the system, all modes are symmetric about the line $n_{2}=0$.

Consider the total force on an inclusion containing $N$ defects
\begin{equation}
F = \sum_{p=0}^{{N-1}}\left(u_{p-1,0} + u_{p+1,0}+2u_{p,1}\right).
\end{equation}
By definition, for a skew-symmetric mode $u_{p,0} = -u_{N-1-p,0}$ and further $u_{p,q} = -u_{N-1-p,q}$.
Hence, for all skew-symmetric modes the inclusion is self-balanced (i.e. $F=0$) and therefore, all skew-symmetric localised modes can be considered as multipole modes.

For the illustrative examples presented later, the eigenvalue problem~\eqref{eq:spectral} will be solved for the unit eigenvectors ($|\mathcal{U}|=1$).

\subsection{Asymptotics}
Here, asymptotics are considered for some particular cases. Asymptotic expansions for an isolated Green's matrix in various configurations have been considered by~\cite{movchan-slepyan} and the approach detailed therein is used here.
\paragraph{In the far field, along the line of defects.}
The case of $n_1\to\infty$, $n_2=0$ and finite $N$ is considered.
Introducing the small parameter $\varepsilon=p/n_1$, the kernel of~\eqref{eq:sgfx1} may be expanded for small $|\varepsilon|\ll1$. In particular,
\begin{equation}
\left(\sqrt{a^2-1}-a\right)^{|n_1-p|}\sim \left(\sqrt{a^2-1}-a\right)^{|n_1|}\left[1-\varepsilon\log\left(\sqrt{a^2-1}-a\right)\right]^{|n_1|},
\end{equation}
it is observed that at large $n_1$ and sufficiently small $N$, the dominant contribution to the integral~\eqref{eq:sgfx1} comes from a small region in the vicinity of $\xi_2=\pi$. Therefore,
\begin{equation}
\begin{split}
\left(\sqrt{a^2-1}-a\right)^{|n_1-p|}& \sim
 \left(\sqrt{c^2-1}-c\right)^{|n_1|}\left[1-\frac{(\pi-\xi_2)^2}{2\sqrt{c^2-1}}\right]^{|n_1|}\\
&
\times\left[1-\varepsilon\log\left(\sqrt{c^2-1}-c\right)+\varepsilon\frac{(\pi-\xi_2)^2}{2\sqrt{c^2-1}}\right]^{|n_1|},
\end{split}
\end{equation}
where $c=\omega^2/2-3$. Thus,
\begin{equation}
\begin{split}
\left(\sqrt{a^2-1}-a\right)&^{|n_1-p|}\\
&\sim \left(\sqrt{c^2-1} - c\right)^{|n_1-p|}\exp\left[-|n_1-p|\frac{(\pi-\xi_2)^2}{2\sqrt{c^2-1}} \right].
\end{split}
\end{equation}
In addition, $1/\sqrt{a^2-1} \sim 1/\sqrt{c^2-1}$.
Hence, for $0<\varepsilon\ll1$ and making use of~\eqref{eq:sgfx1}
\begin{equation}
g(n_1,0,p;\omega) \sim \frac{\left(\sqrt{c^2-1} - c\right)^{|n_1-p|}}{2\pi\sqrt{c^2-1}}\int\limits_{\pi-\varepsilon}^\pi\exp\left[-|n_1-p|\frac{(\pi-\xi_2)^2}{2\sqrt{c^2-1}} \right] \D\xi_2.
\end{equation}
Making the substitution $x = (\pi-\xi_2)\sqrt{|n_1-p|/2\sqrt{c^2-1}}$, and performing the resulting integration yields
\begin{equation}
g(n_1,0,p;\omega) \sim \frac{\left(\sqrt{c^2-1} - c\right)^{|n_1-p|}}{\sqrt{8\pi\sqrt{c^2-1}}}\frac{1}{\sqrt{|n_1-p|}} \quad \text{as}\quad n_1\to\infty.
\label{eq:gp-parallel}
\end{equation}
Thus from~\eqref{eq:field}, the physical field has the following approximate representation for $n_1\to\infty$
\begin{equation}
\label{eq:on-force-line}
u_{n_{1},0}(\omega) \sim (1-r)\omega^2\sum_{p=0}^{N-1}\frac{\left(\sqrt{c^2-1} - c\right)^{|n_1-p|}}{\sqrt{8\pi\sqrt{c^2-1}}}\frac{u_{p,0}(\omega)}{\sqrt{|n_1-p|}},
\end{equation}
where $u_{p,0}(\omega)$ should be determined from~\eqref{eq:spectral}.
It is observed that when $N=1$ equation~\eqref{eq:on-force-line} is consistent with equation (4.17) of~\cite{movchan-slepyan} up to a change in sign.

\paragraph{In the far field, perpendicular to the line of defects.}
Here, the case considered is $n_1=p^{\prime}$, $n_2\to\infty$ with $N$ and $p^{\prime}$ finite.
The method used here follows the same general procedure as in the previous case.
However in this case, the kernel is oscillatory and is therefore approximated as a product of decaying and oscillatory functions.

For sufficiently small $|p^{\prime}-p|$ and large  $n_2$, the non-oscillatory part of the integrand in~\eqref{eq:sgfx1} is approximated as before, leading to
\begin{equation}
\begin{split}
g(p^\prime,n_2,p;\omega) \sim & \frac{\left(\sqrt{c^2-1} - c\right)^{|n_2|}}{2\pi\sqrt{c^2-1}}\\
& \qquad\qquad\times\int\limits_{\pi-\varepsilon}^\pi\exp\left[-|n_2|\frac{(\pi-\xi_1)^2}{2\sqrt{c^2-1}} \right]
\cos{([p^{\prime}-p]\xi_1)} \D\xi_1.
\end{split}
\end{equation}
Making a similar change of variable, $x = (\pi-\xi_1)\sqrt{|n_2|/2\sqrt{c^2-1}}$, and integrating, it is found that
\begin{equation}
\begin{split}
g(p^{\prime},n_2,p;\omega) \sim  (-1)^{(p^{\prime}-p)}&\frac{\left(\sqrt{c^2-1} - c\right)^{|n_2|}}{\sqrt{8\pi\sqrt{c^2-1}}}\frac{1}{\sqrt{|n_2|}} \\
&\qquad\qquad\times\exp\left[-(p^{\prime}-p)^2\frac{\sqrt{c^2-1}}{2|n_2|}\right].
\label{eq:gp-perp}
\end{split}
\end{equation}
Hence, for $n_2\to\infty$ the physical field in~\eqref{eq:field} may be approximated by
\begin{equation}
\label{eq:perp-force-line}
\begin{split}
u_{p^{\prime},n_{2}}(\omega)  \sim (1-r)\omega^2&\frac{\left(\sqrt{c^2-1} - c\right)^{|n_2|}}{\sqrt{8\pi\sqrt{c^2-1}}}\\
&\times\sum_{p=0}^{N-1}(-1)^{(p^{\prime}-p)}\exp\left[-(p^{\prime}-p)^2\frac{\sqrt{c^2-1}}{2|n_2|}\right]\frac{u_{p,0}(\omega)}{\sqrt{|n_2|}}.
\end{split} 
\end{equation}
It is observed that for $N=1$ and $p^{\prime}=p$, the above equation~\eqref{eq:perp-force-line} is consistent with equation (4.17) of~\cite{movchan-slepyan} up to a change in sign. Moreover, for the case of $p^{\prime}=p$,~\eqref{eq:perp-force-line} reduces to~\eqref{eq:on-force-line}.

\subsection{In the vicinity of the band edge}
The representations of Green's matrix~\eqref{eq:sgfx1}-\eqref{eq:g-hypergeom} presented above are valid in the stop band.
However, given that the hypergeometric function in the representation~\eqref{eq:g-hypergeom} is zero balanced, that is, the sum of the bottom parameters minus the sum of the top parameters vanishes: $2(b_{1}+b_{2})- 1 -2(a_{1}+a_{2}) = 0$, the stop band Green's matrix can be extended to the boundary of the pass band by analytic continuation\footnote{
Indeed, for any integer balanced hypergeometric function $_{q+1}\mathrm{F}_q$ there exists an analytic continuation to the boundary of the unit disk (see~\cite{burhring}, among others, for details).}.
In particular, the analytical continuation of the function~\eqref{eq:g-hypergeom} has the form
\begin{equation}
\label{anal-cont}
\begin{split}
g(\vec{n},p;\omega) & =  \frac{(-4)^{m+n_2}}{\pi(2\alpha)^{1+m+n_2}}\sum_{j=0}^\infty\left(\frac{\left([1+m+n_2]/2\right)_j}{j!}\right)^2\left(1-\frac{4}{\alpha^2}\right)^j\\
&\quad\times\left\{\sum_{k=0}^j\frac{(-j)_k}{\left\{\left([1+m+n_2]/2\right)_j\right\}^2}\mathfrak{F}(m,n_2,k)\left[
\phantom{\frac{1}{1^1}}\hspace{-3ex}\psi(1+j-k) \right.\right. \\
&\qquad\left.\left. + \psi(1+j)-\psi\left(\frac{1+m+n_2}{2}+j\right)- \log\left(1-\frac{4}{\alpha^2}\right)\right] \right.\\
&\qquad\quad\left.+
(-1)^j(j)!\sum_{k=j+1}^\infty\frac{(k-j-1)!}{\left\{([1+m+n_2]/2)_k\right\}^2}\mathfrak{F}(m,n_2,k)\right\}
\end{split}
\end{equation}
where the reader is reminded that $m=n_1-p$, $(\cdot)_j$ is the Pochhammer symbol, $\psi(x)$ is the Digamma function,  and
\begin{equation}
\mathfrak{F}(m,n,k) = \frac{(m)_k(n)_k}{k!}\pFq{3}{2}{(m+n_2)/2,(m+n_2)/2,-k}{m,n}{1}.
\label{eq:f-frak}
\end{equation}
The symbol $_{p}\mathrm{F}_q$[\ldots] denotes the generalised hypergeometric function, which is related to the regularised generalised hypergeometric function thus:
$$_{p}\mathrm{F}_{q}[a_{1},\ldots,a_{p};b_{1},\ldots b_{q};z] =\{\Gamma(b_{1})\ldots\Gamma(b_{q})\}\; _{p}\mathbf{F}_{q}[a_{1},\ldots,a_{p};b_{1},\ldots b_{q};z]$$.
In this case, the continuation~\eqref{anal-cont} holds for $\alpha^2\geq4$, which in terms of frequency corresponds to $\omega^2\geq8$.
It is emphasised that in this section, the term ``\emph{vicinity of the band edge}'' refers to a small interval $8\leq\omega^2<8+\varepsilon$, where $0<\varepsilon\ll1$.

Hence, choosing $j=0$ yields the leading order behaviour of~\eqref{eq:g-hypergeom} as  $\alpha^{2}\to4^+$ ($\omega^2\to8^+$), that is, as $\omega$ approaches the boundary of the pass band from the stop band:
\begin{equation}
\begin{split}
g(\vec{n},p;\omega) \sim
 \frac{(-4)^{m+n_2}}{\pi(2\alpha)^{1+m+n_2}}&\left\{
\left[-2\gamma-\psi\left(\frac{1+m+n_2}{2}\right)-\log\left(1-\frac{4}{\alpha^2}\right)\right]\right.\\
&\left.
+\sum_{k=1}^\infty\frac{(k-1)!}{\left\{([1+m+n_2]/2)_k\right\}^2}\mathfrak{F}(m,n_2,k)\right\},
\end{split}
\label{eq:band-edge}
\end{equation}
where $\gamma$ is the Euler-Mascheroni constant.
Alternative representations of the leading order continuations for general zero-balanced $_{q+1}\mathrm{F}_q$ were derived by~\cite{saigo}.
Since $k>0$, the series representation of the hypergeometric function in~\eqref{eq:f-frak} has a finite number of terms and therefore may be computed exactly.
The convergence condition for the infinite sums in~\eqref{anal-cont} and~\eqref{eq:band-edge} is $2+m+n_2+j>0$, and is automatically satisfied since it was assumed (without loss of generality) at the outset that $m\geq0$ and $n_2\geq0$.

The asymptotic expression~\eqref{eq:band-edge} is particularly interesting as it elucidates the nature of the singularity of the lattice Green's matrix at the band edge.
In particular, the asymptotic representation~\eqref{eq:band-edge} captures the logarithmic singularity as $\omega^2\to8^+$.
This logarithmically singular behaviour near the band edge is not obvious from the original representations presented earlier (cf. equations ~\eqref{eq:sgfx1}-\eqref{eq:g-hypergeom}).

\begin{SCfigure}[1]
\includegraphics[width=0.5\linewidth]{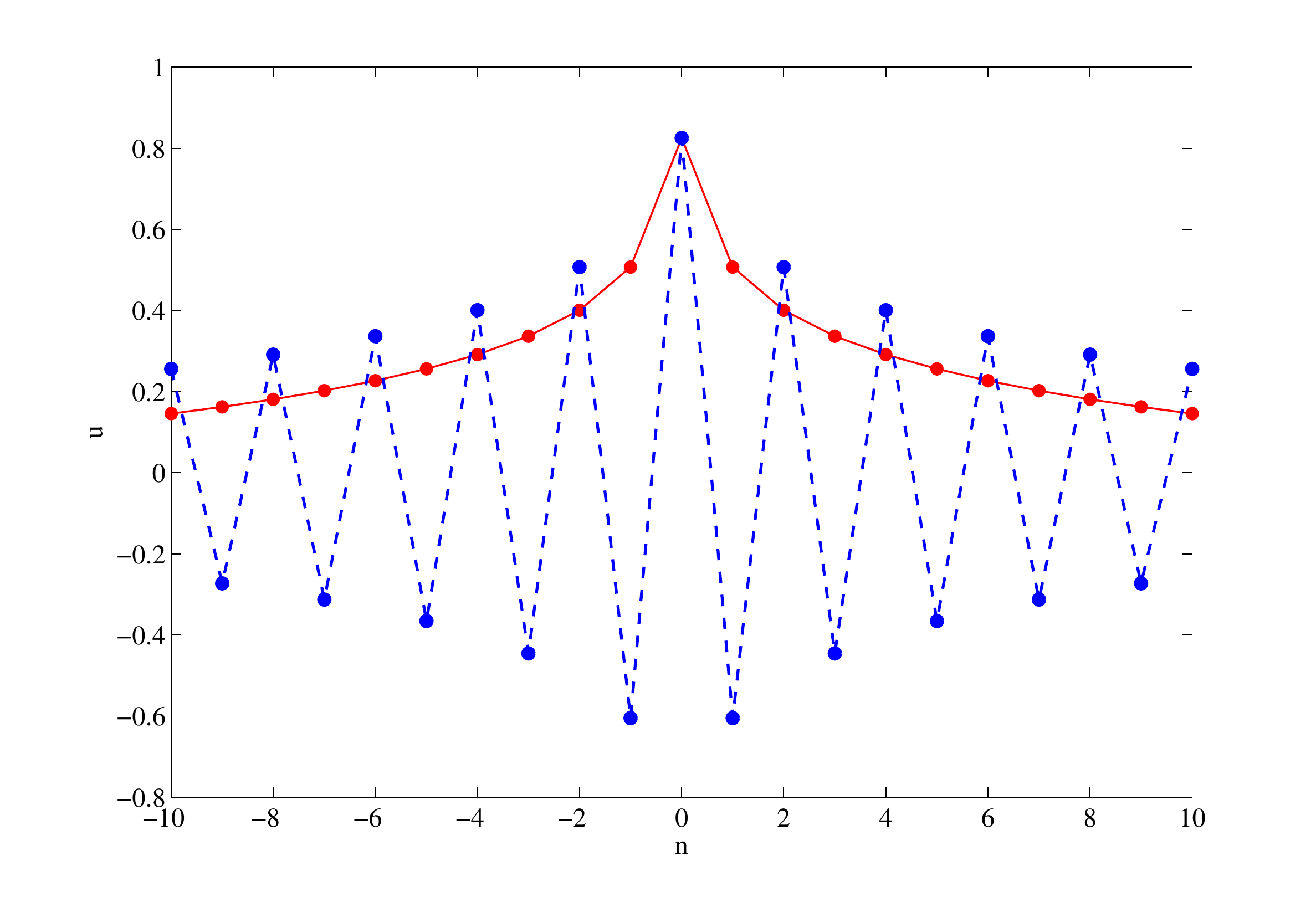}
\caption{\label{fig:band-edge-gf}
The solid curve shows the asymptotic expression for the displacement field along the diagonal ($n_1=n_2$ with $p=0$) in the vicinity of the band edge (cf. equation~\eqref{eq:diagonal-band-edge}). The dashed curve shows the corresponding asymptotic expression for the field along the bond line (cf. equation~\eqref{eq:bond-line-band-edge}). The frequency chosen was $\omega=2.829$.
}
\end{SCfigure}

For some particular cases, equation~\eqref{eq:band-edge} reduces to the following simplified forms.
\begin{subequations}
\label{eq:band-edge-spec}
Along the rays $m=0$ (i.e. $n_1=p$) or replacing $n_{2}$ by $n_{1}-p$, along $n_{2}=0$:
\begin{equation}
g(p,n_2,p;\omega) \sim
\frac{(-4)^{1+n_2}}{\pi(2\alpha)^{1+n_2}}\left[2\gamma+\psi\left(\frac{1+n_2}{2}\right)+\log\left(1-\frac{4}{\alpha^2}\right)\right]
\triangleq \tilde{g}^\text{(bond)}(n_{2};\omega),
\label{eq:bond-line-band-edge}
\end{equation}
and along the diagonal rays $m=n_{2}$ (i.e. $n_1=n_2+p$):
\begin{equation}
g(n_{1},m,p;\omega) \sim
-\frac{16^{m}}{\pi(2\alpha)^{1+2m}}\left[2\gamma+\psi\left(\frac{1}{2}+m\right)+\log\left(1-\frac{4}{\alpha^2}\right)\right]
\triangleq \tilde{g}^\text{(diag)}(m;\omega),
\label{eq:diagonal-band-edge}
\end{equation}
\end{subequations}
where the reader is reminded that $m=n_{1}-p$.
The Digamma function grows logarithmically as $m\to\infty$ and the term $2\gamma+\psi(1/2+m)$ is strictly positive for $m>0$.
Therefore, for sufficiently small $m$ the bracketed term in equations~\eqref{eq:band-edge-spec} is negative in the neighbourhood of $\alpha=2$.
Hence, in the vicinity of the band edge, the stop band Green's matrix exhibits fundamentally different behaviour along the bond lines compared with the diagonal rays.
In particular, along the bond lines the masses will oscillate out of phase, whereas for the diagonal ray lines the masses will oscillate in phase, as illustrated in figure~\ref{fig:band-edge-gf}.
In the far field, equations~\eqref{eq:band-edge-spec} further reduce to
\begin{subequations}
\label{eq:band-edge-spec-far field}
\begin{equation}
g(p,n_2,p;\omega) \sim 
\frac{(-4)^{1+n_2}}{\pi(2\alpha)^{1+n_2}}\left[2\gamma+\log\left(\frac{n_2}{2}\right)+\log\left(1-\frac{4}{\alpha^2}\right)\right]\text{ as } n_2\to\infty,
\label{eq:bond-line-band-edge-far-field}
\end{equation}
\begin{equation}
g(m,m,p;\omega) \sim
-\frac{16^{m}}{\pi(2\alpha)^{1+2m}}\left[2\gamma+\log m+\log\left(1-\frac{4}{\alpha^2}\right)\right]
\text{ as } m\to\infty.
\label{eq:diagonal-band-edge-far-field}
\end{equation}
\end{subequations}
Using equations~\eqref{eq:field} and~\eqref{eq:band-edge-spec} the anti-plane displacement for a lattice with $N$ defects has the following asymptotic representation in the vicinity of the band edge
\begin{subequations}
\begin{equation}
u_{n_{1},0}(\omega) \sim (1-r)\omega^2\sum_{p=0}^{N-1}u_{p,0}\tilde{g}^\text{(bond)}(n_{1}-p;\omega),\;\text{as}\;\omega^{2}\to8^{+},
\label{eq:band-edge-field-n2-0}
\end{equation}
\begin{equation}
u_{n_{1},n_2-p}(\omega) \sim (1-r)\omega^2\sum_{p=0}^{N-1}u_{p,0}\tilde{g}^\text{(diag)}(n_{1}-p;\omega),\;\text{as}\;\omega^{2}\to8^{+},
\label{eq:band-edge-field-n2-m}
\end{equation}
\end{subequations}
along the rays $n_{2}=0$ and $n_{2} = n_{1}-p$ respectively.

\section{Illustrative examples}\label{sec:illustex}
\begin{figure}
\centering
\subfigure[\label{fig:c-p0}A single defect ($N=1$)]{
\includegraphics[width=0.3\linewidth]{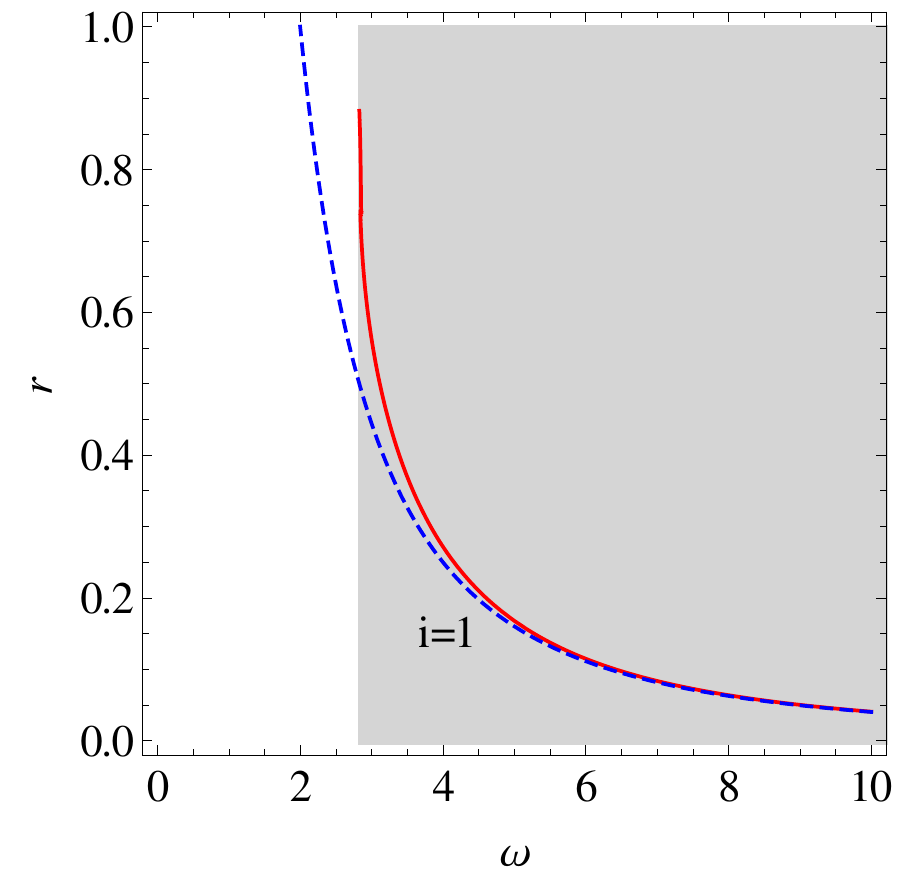}
}
\subfigure[\label{fig:c-p1}A pair of defects ($N=2$)]{
\includegraphics[width=0.3\linewidth]{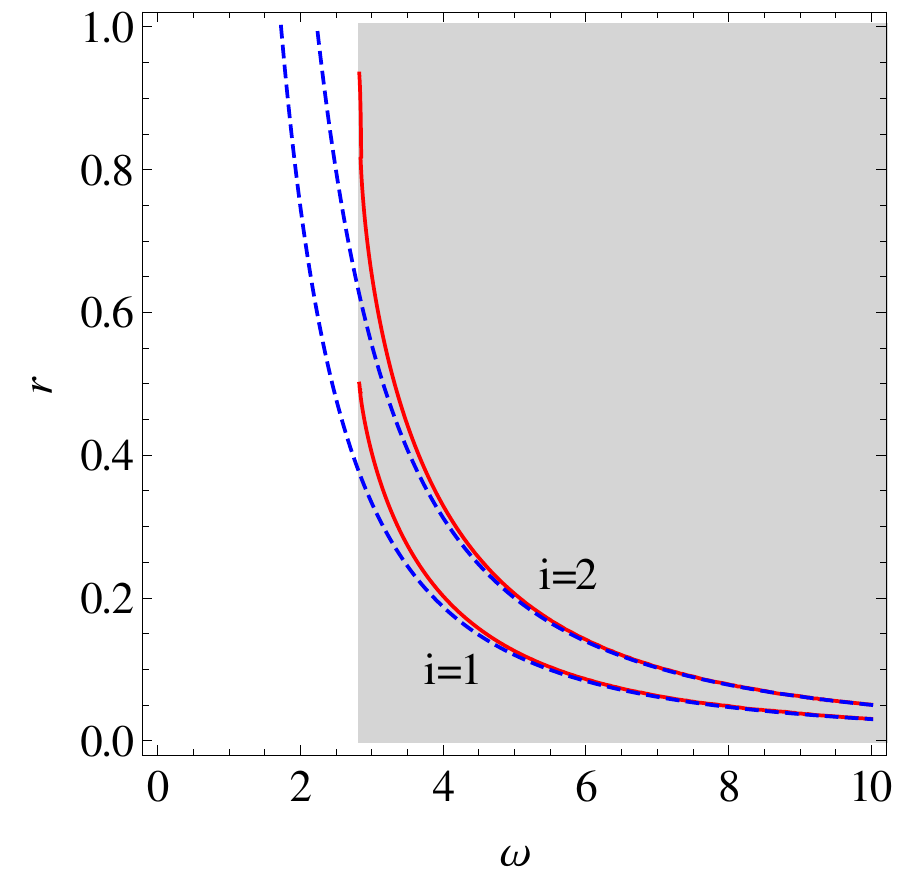}
}
\subfigure[\label{fig:c-p2}A triplet of defects ($N=3$)]{
\includegraphics[width=0.3\linewidth]{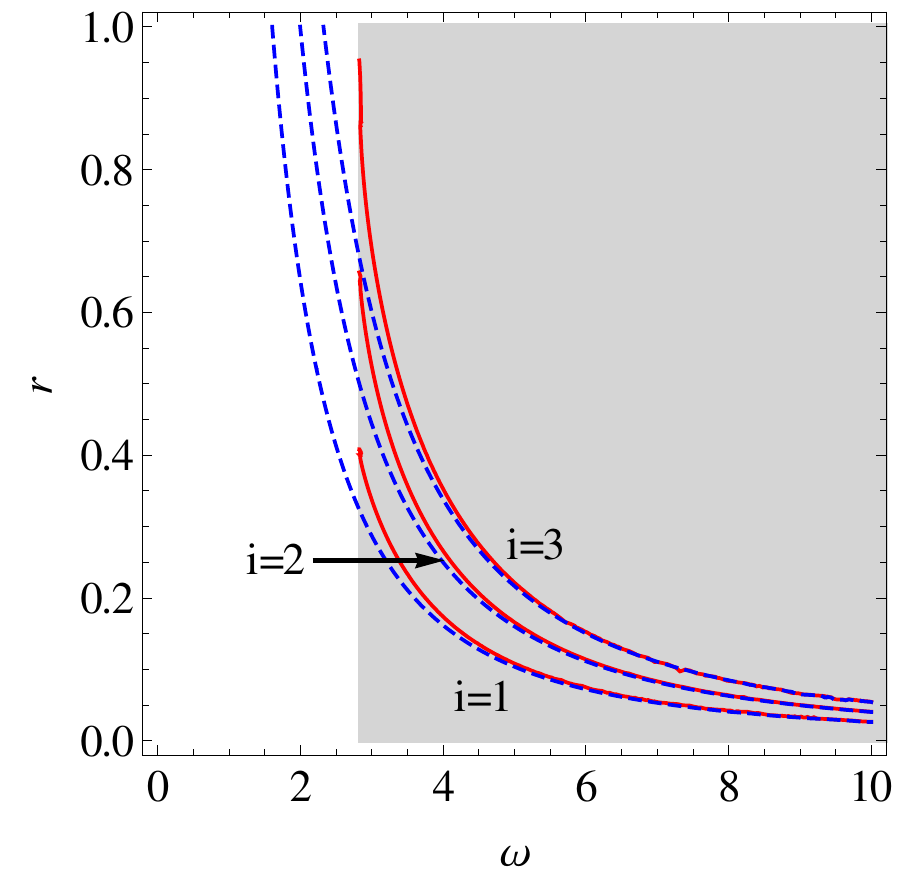}
}
\caption{\label{fig:phase-diagrams}
The solid curves show the $i^\text{th}$ solution, $r_{N,i}(\omega)$, of the solvability condition~\eqref{eq:trans} for a system of $N$ defects embedded in the square lattice.
The shaded region ($\omega^2>8$) indicates the stop band of the ambient lattice.
The dashed curves show the corresponding $i^\text{th}$ solution, $r^{*}_{N,i}(\omega)$, of the solvability condition for an isolated system of $N$ defects (cf. equation~\eqref{eq:spec-line}).
}
\end{figure}

Several particular cases are considered here corresponding to relatively short defects with $N\in[1,3]$.
The solid curves in figure~\ref{fig:phase-diagrams} show the $i^\text{th}$ solution, $r_{N,i}(\omega)$, of the solvability condition~\eqref{eq:trans} for a line of $N$ defects.
The shaded region indicates the stop band ($\omega^2>8$) of the ambient lattice.
For frequencies in this region, waves in the ambient lattice will decay exponentially away from the defect or source. 
It is interesting to note that in contrast to the 1D  and 3D cases (see for example,~\cite{maradudin1965}) the image of $r_{N,N}(\omega)$, indicated by the solid curves in figure~\ref{fig:phase-diagrams}, is $(0,1)$.
In other words, a localised defect mode can be initiated by creating a defect in the lattice by removing any amount of mass from one or more nodes. In 1D and 3D lattices, there is some upper bound on the ratio of the mass of the defect to the ambient lattice such that a localised mode can be initiated.
As $r\to1$, that is, the lattice approaches a homogeneous lattice, the frequency of the localised mode approaches the band edge ($\omega^{2}\to8^+$).
It is also observed that for $N>1$, the solid curves intersect the band edge at several distinct values of $r$.
This suggests that for a given number of defects, there exists a maximum value of $r$ below which all possible localised eigenmodes may be initiated.
Above this value of $r$ it is only possible to initiate a subset of the possible eigenmodes with the lower frequency eigenmodes being filtered out.
In all cases, the highest frequency eigenmode persists for all possible values of $r$ on $(0,1)$.
For fixed $\omega$, the solvability condition~\eqref{eq:trans} for a system of $N$ defects is a polynomial, of at most degree $N$, in $r$.
Therefore, there exist no more than $N$ solutions for a given frequency $\omega$.

The dashed curves correspond to the problem of an isolated chain of $N$ particles of non-dimensional mass $r^*$, connected by springs to two nearest neighbours and surrounded by rigid foundations.
For such a problem, the out-of-plane displacement of mass $n\in\mathbb{Z}$ satisfies
\begin{equation}
\mathcal{L}[
v_0, v_1, \cdots, v_{N-1}
]^\mathrm{T}
= 0,
\label{eq:spec-line}
\end{equation}
where the matrix $\mathcal{L}$ has elements
\begin{equation}
[\mathcal{L}]_{ij} = (r^*\omega^2-4)\delta_{ij} + \delta_{i-1,j}+\delta_{i,j-1}.
\label{eq:L-line}
\end{equation}
The dashed curves in figure~\ref{fig:phase-diagrams} represent the solutions $r^{*}_{N,i}(\omega)$ of the solvability condition: $\det\mathcal{L} = 0$.
It is observed that as $\omega\to\infty$, the dashed curves approach the solid curves from below.

\subsection{A single defect}
For the case of a single defect located at the origin, the quantity $\mathcal{G}$ in~\eqref{eq:spectral} is a scalar:
\begin{equation}
\mathcal{G}(\omega) = \frac{1}{\alpha\pi}K\left(\frac{4}{\alpha^2}\right),
\label{eq:G11}
\end{equation}
where $K(x)$ is the complete elliptical integral of the first kind.
The solvability condition may be written as
\begin{equation}
r_{1,1} = 1+\pi\left(\frac{2}{\omega^2}-\frac{1}{2}\right)\left[K\left(\frac{16}{(\omega^2-4)^2}\right)\right]^{-1},
\end{equation}
which has the leading order asymptotic representation
\begin{equation}
r_{1,1} \sim \frac{4}{\omega^2}, \qquad\text{as}\qquad \omega\to\infty.
\label{eq:r0-asymp}
\end{equation}
It is observed that the solvability condition for equation~\eqref{eq:spec-line} with $N=1$ agrees precisely with the leading order high frequency asymptotic expansion.
Hence, the observed coalescence of the solid and dashed curves in figure~\ref{fig:c-p0}.

\begin{figure}[h!tb]
\centering
\subfigure[\label{fig:p0-m1}]{
\includegraphics[width=0.35\linewidth]{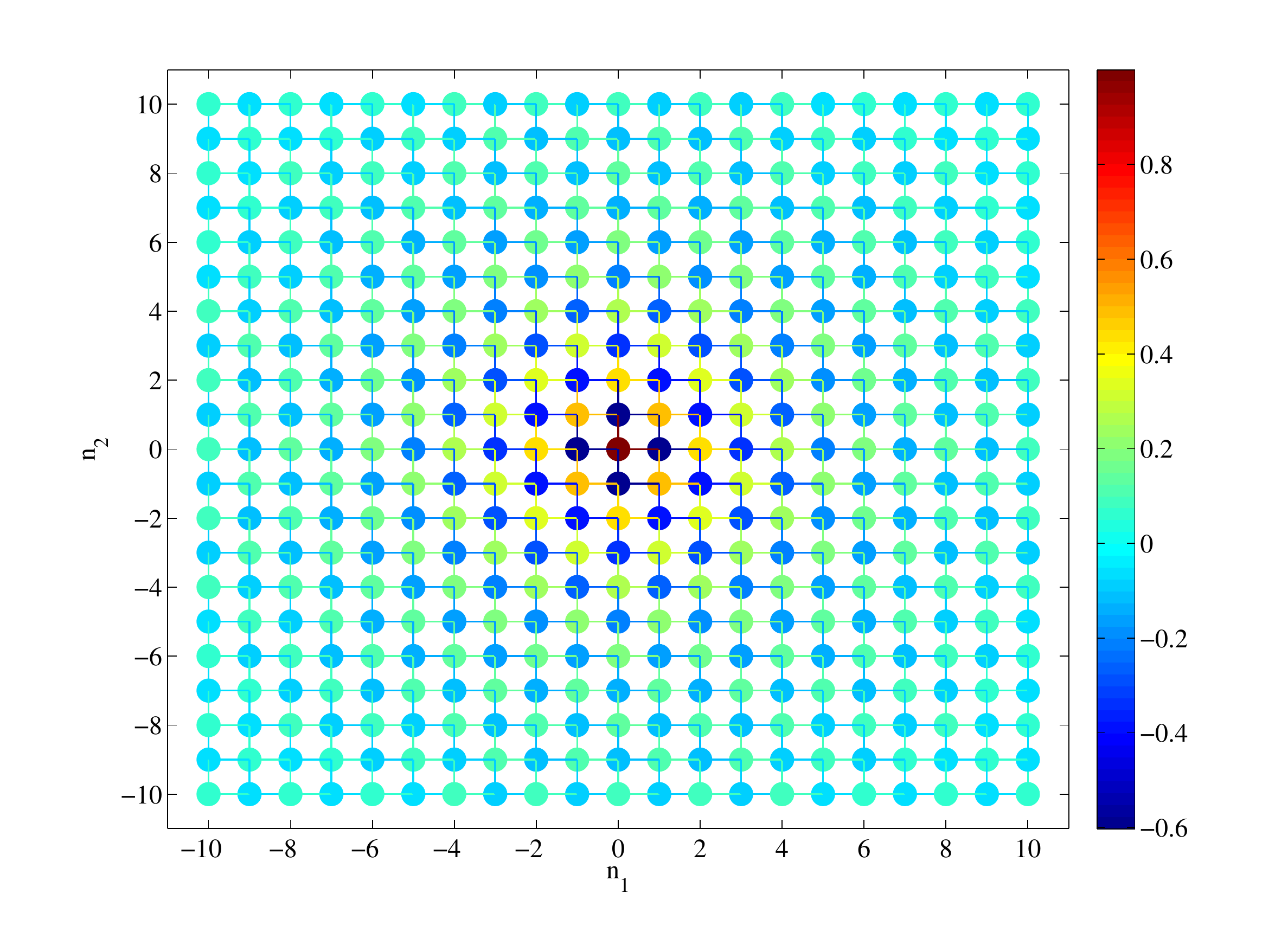}
}\qquad
\subfigure[\label{fig:p0-m1-n2-0}]{
\includegraphics[width=0.35\linewidth]{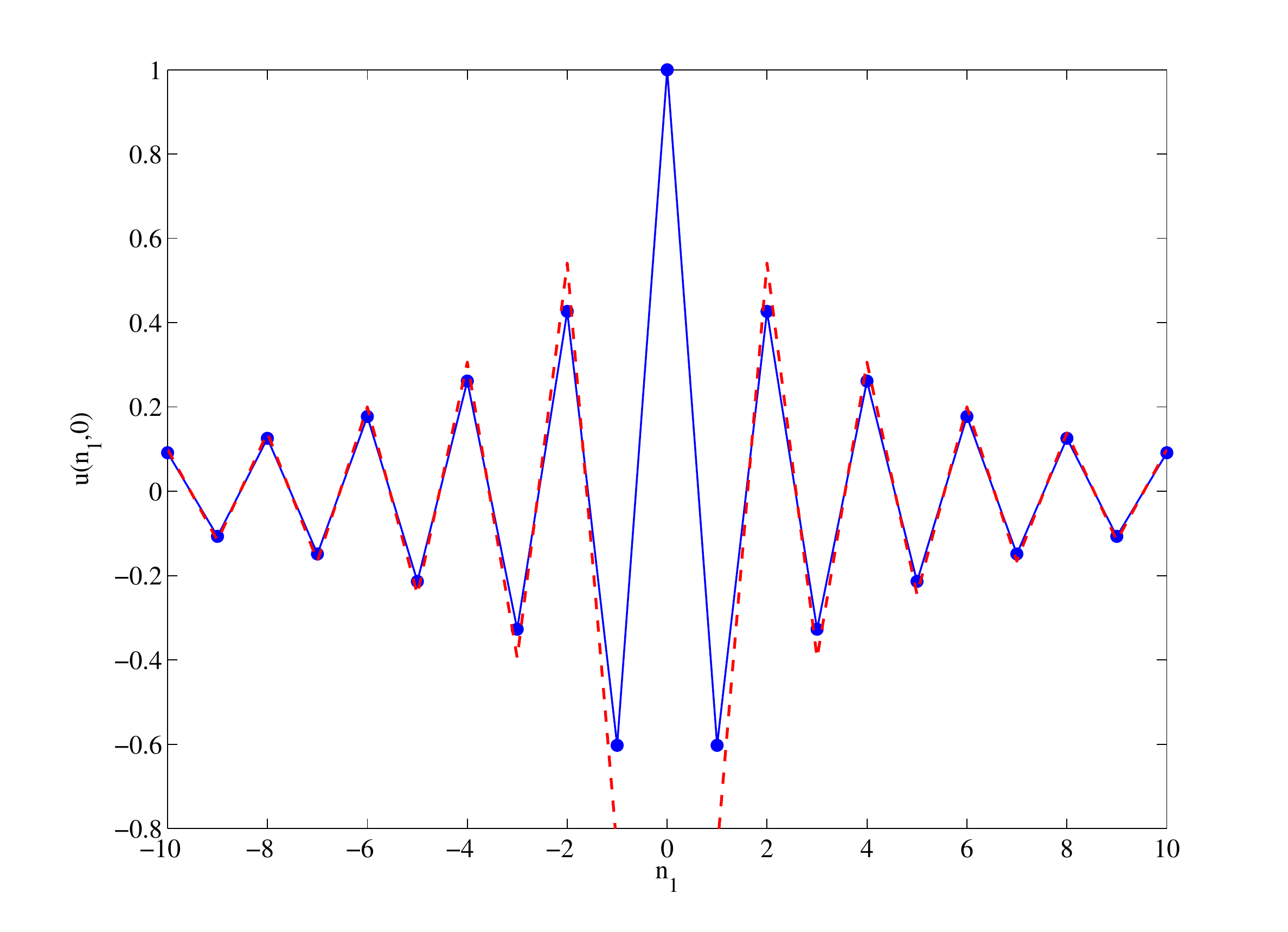}
}\qquad
\subfigure[\label{fig:p0-m1-band-edge-diag}]{
\includegraphics[width=0.35\linewidth]{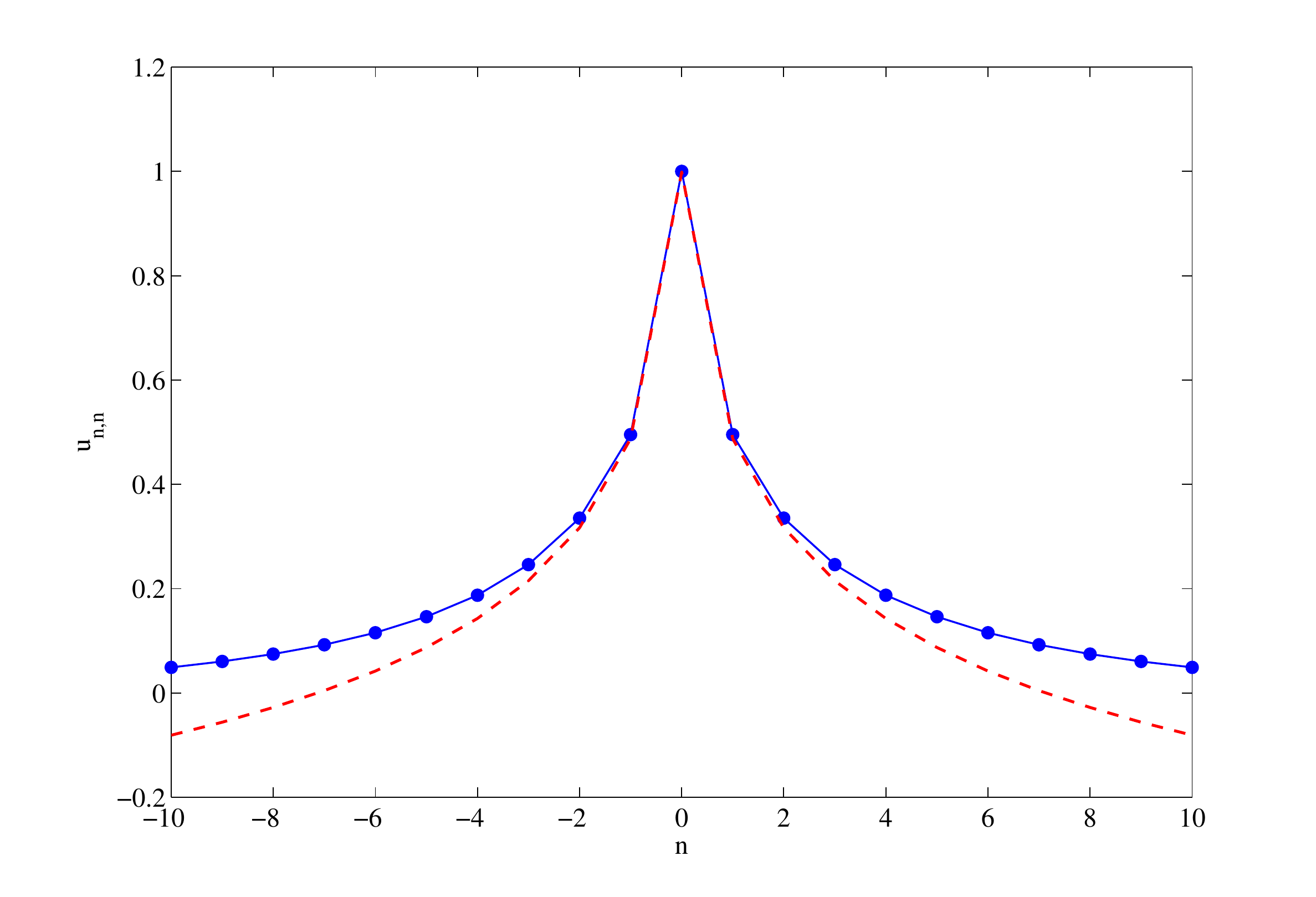}
}\qquad
\subfigure[\label{fig:p0-m1-band-edge-n2-0}]{
\includegraphics[width=0.35\linewidth]{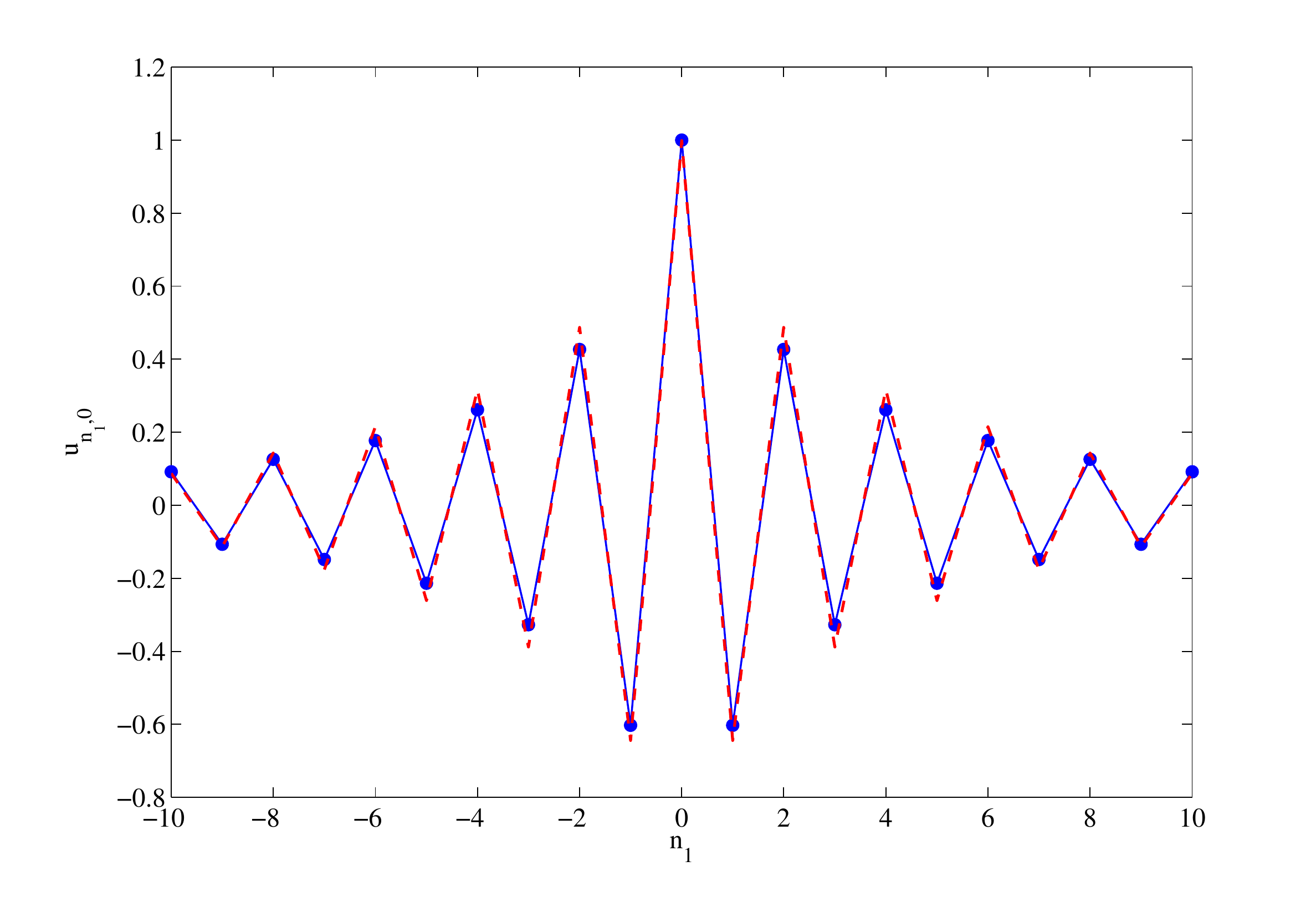}
}
\caption{\label{fig:p0}
(a) The localised defect mode for a single defect with $r=0.8$ and $\omega=2.83$.
(b) The solid curve is the out-of-plane displacement along the line $n_2=0$ and the dashed curve is the asymptotic expansion for $n_1\to\infty$ (cf.~\eqref{eq:on-force-line}).
(c) The out-of-plane displacement along the line $n_{1}=n_{2}$ (solid curve) with the corresponding asymptotic expansion~\eqref{eq:diagonal-band-edge} for the  band edge (dashed curve).
(d) As for (b), but the dashed curve represents the band edge expansion along $n_{2}=0$ (cf. equation~\eqref{eq:band-edge-field-n2-0}).}
\end{figure}
The localised defect mode is shown in figure~\ref{fig:p0-m1}, together with field along the line $n_2=0$ and the associated asymptotic field as $n_1\to\infty$ in figure~\ref{fig:p0-m1-n2-0}.
Figures~\ref{fig:p0-m1-band-edge-diag} and~\ref{fig:p0-m1-band-edge-n2-0} show the field (solid line) and the
band edge asymptotics (dashed) for a value of $\alpha=2.006$.
The asymptotic expansions show good agreement with the computed field, even for the far field asymptotics in the neighbourhood of the defect.

\subsection{A pair of defects}
In the case of a pair of defects, $\mathcal{G}(\omega)$ is a $2\times2$ matrix with the diagonal elements given by~\eqref{eq:G11}. The off-diagonal elements have the form
\begin{equation}
\left[\mathcal{G}(w)\right]_{12}  = \frac{1}{4} -\frac{1}{2\pi}K\left(\frac{4}{\alpha^2}\right).
\label{eq:G12}
\end{equation}
The solutions of the solvability condition are
\begin{align}
\label{eq:p1-r1}
r_{2,1} & =1-\frac{4 \pi  (\omega^2-4)}{\pi \omega^2 (\omega^2-4)-2 \omega^2(\omega^2-8) K\left(\dfrac{16}{(\omega^2-4)^2}\right)},
\\
r_{2,2} & =1+\frac{4 \pi(\omega^2-4)}{\pi  \omega^2 (\omega^2-4)-2 \omega^4 K\left(\dfrac{16}{(\omega^2-4)^2}\right)},
\end{align}
whence the leading order high frequency asymptotic expansions are
\begin{equation}
r_{2,1} \sim \frac{3}{\omega^2}\qquad\text{and}\qquad
r_{2,2} \sim \frac{5}{\omega^2}
\quad\text{as}\quad\omega\to\infty,
\end{equation}
which again, agree precisely with the solvability condition of the isolated system~\eqref{eq:spec-line} for $N=2$.
Hence, the observed coalescence of the solid and dashed curves in figure~\ref{fig:phase-diagrams}.

\begin{figure}[h!tb]
\centering
\subfigure[\label{fig:p1-m1}
Symmetric mode at $\omega=2.84$]{
\includegraphics[width=0.35\linewidth]{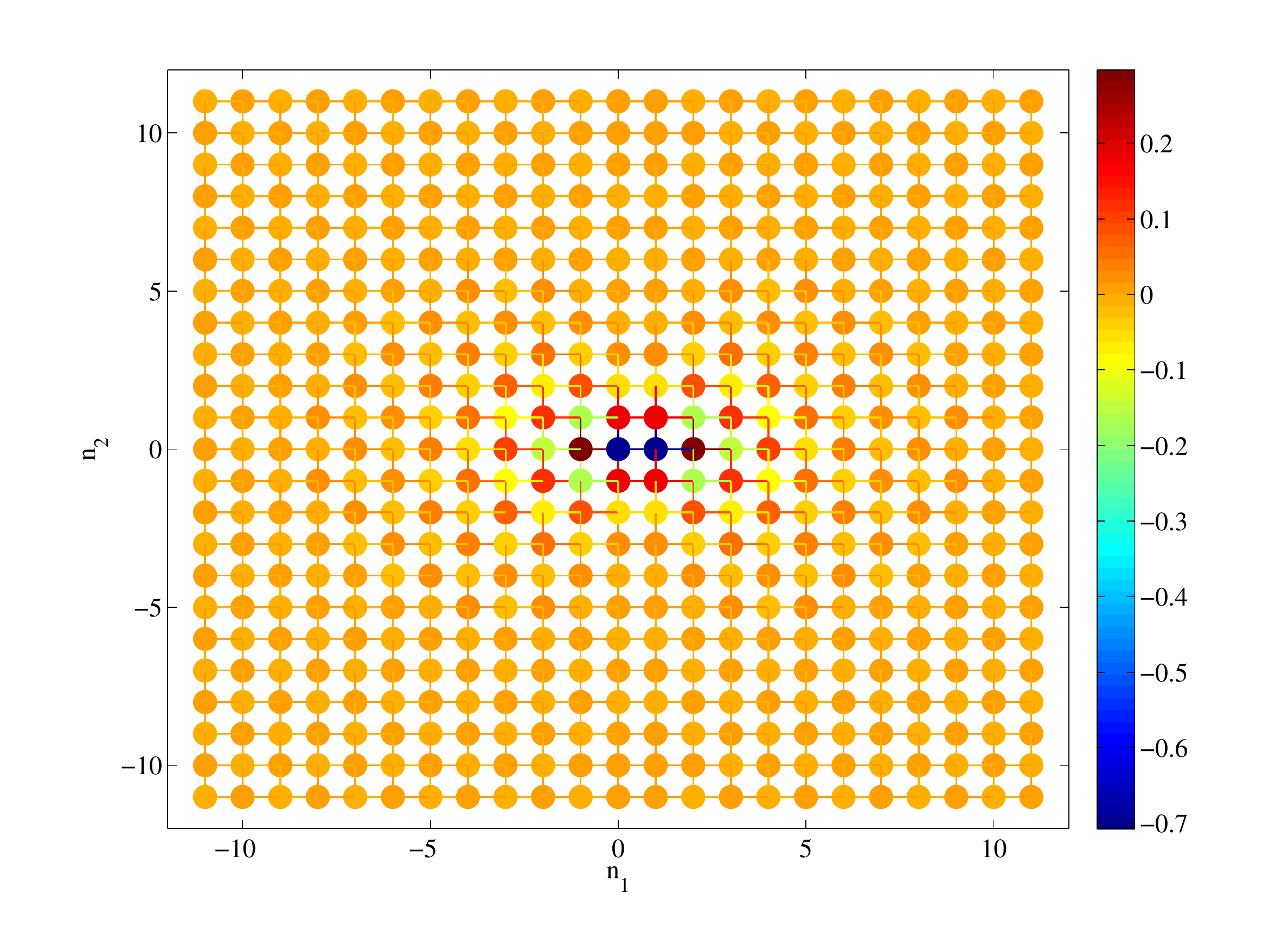}
}\qquad
\subfigure[\label{fig:p1-m2}
Skew-symmetric mode at $\omega=3.35$]{
\includegraphics[width=0.35\linewidth]{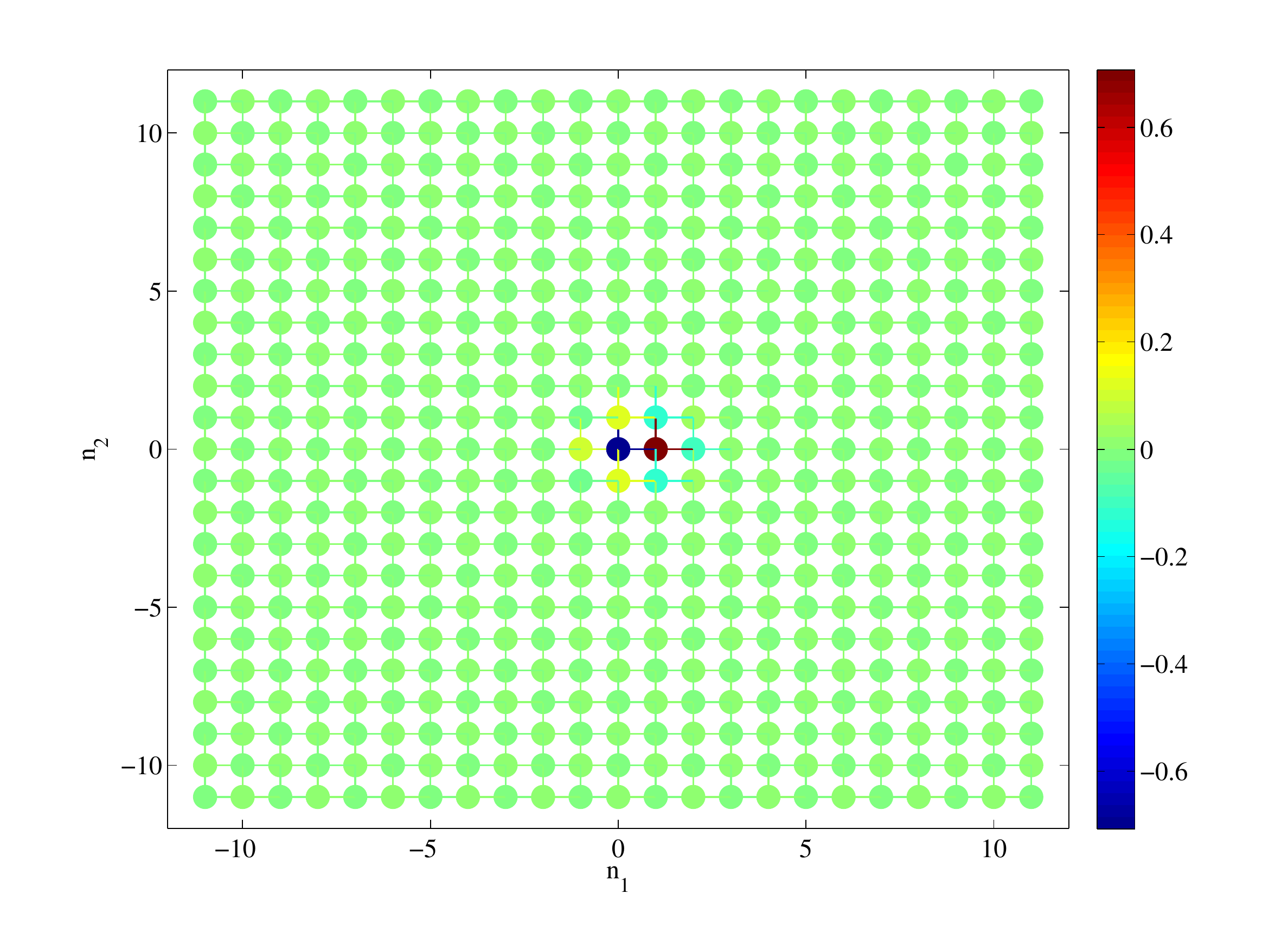}
}\qquad
\subfigure[\label{fig:p1-m1-n2-0}
The field along the line $n_2=0$ for the symmetric mode]{
\includegraphics[width=0.35\linewidth]{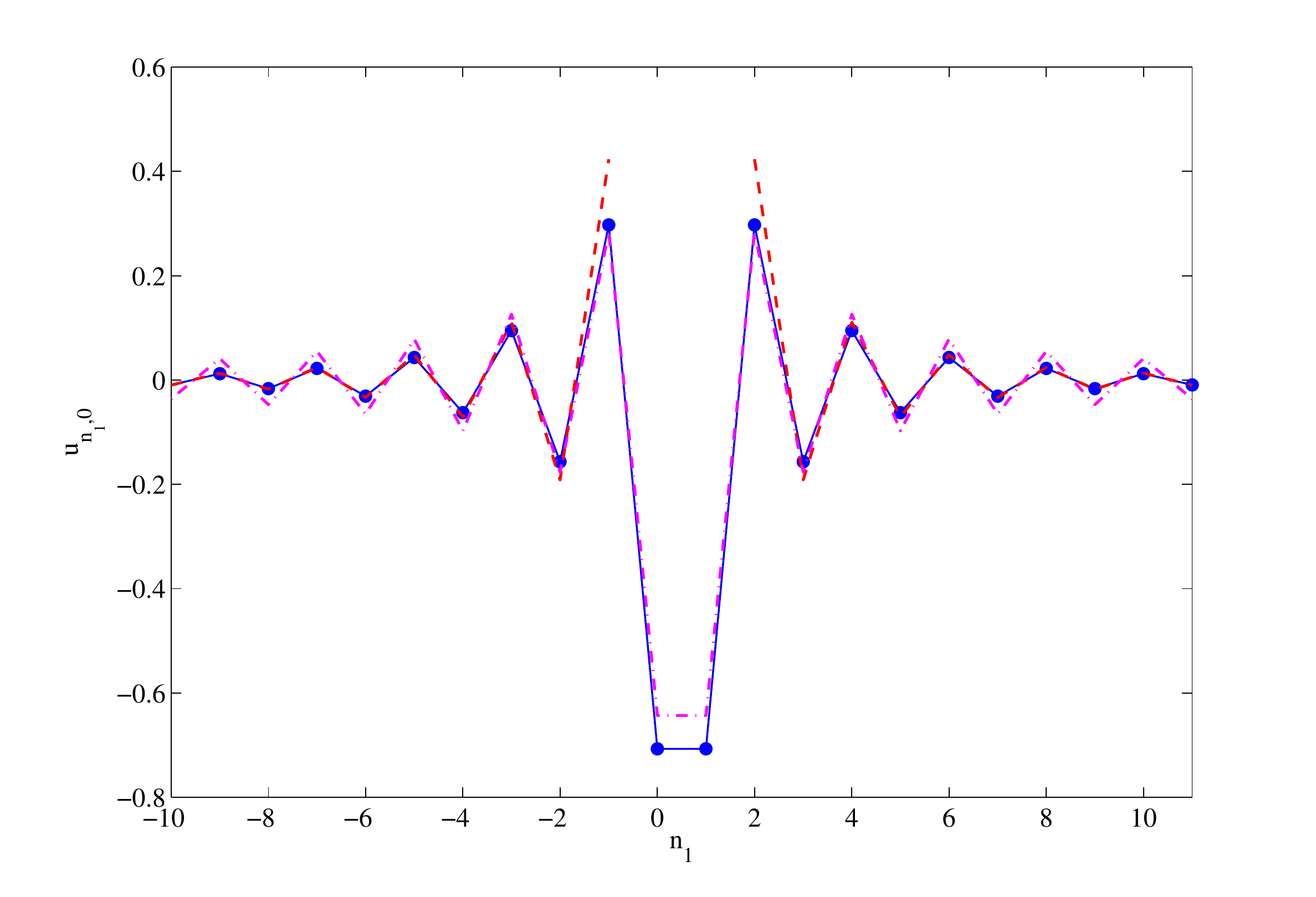}
}\qquad
\subfigure[\label{fig:p1-m2-n2-0}
The field along the line $n_2=0$ for the skew-symmetric mode]{
\includegraphics[width=0.35\linewidth]{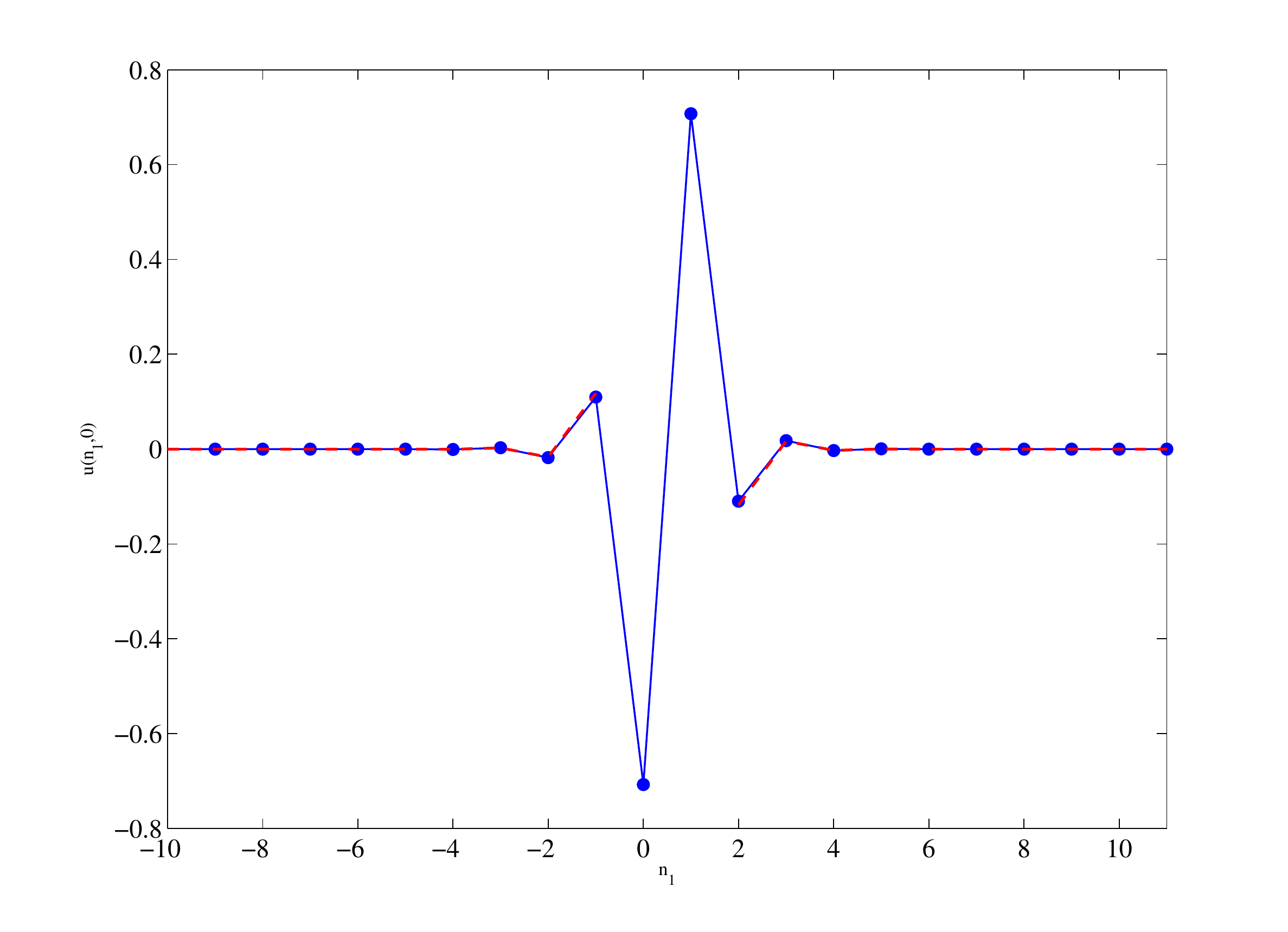}
}\qquad
\subfigure[\label{fig:p1-m1-n1-0}
The field along the line $n_1=0$ for the symmetric mode]{
\includegraphics[width=0.35\linewidth]{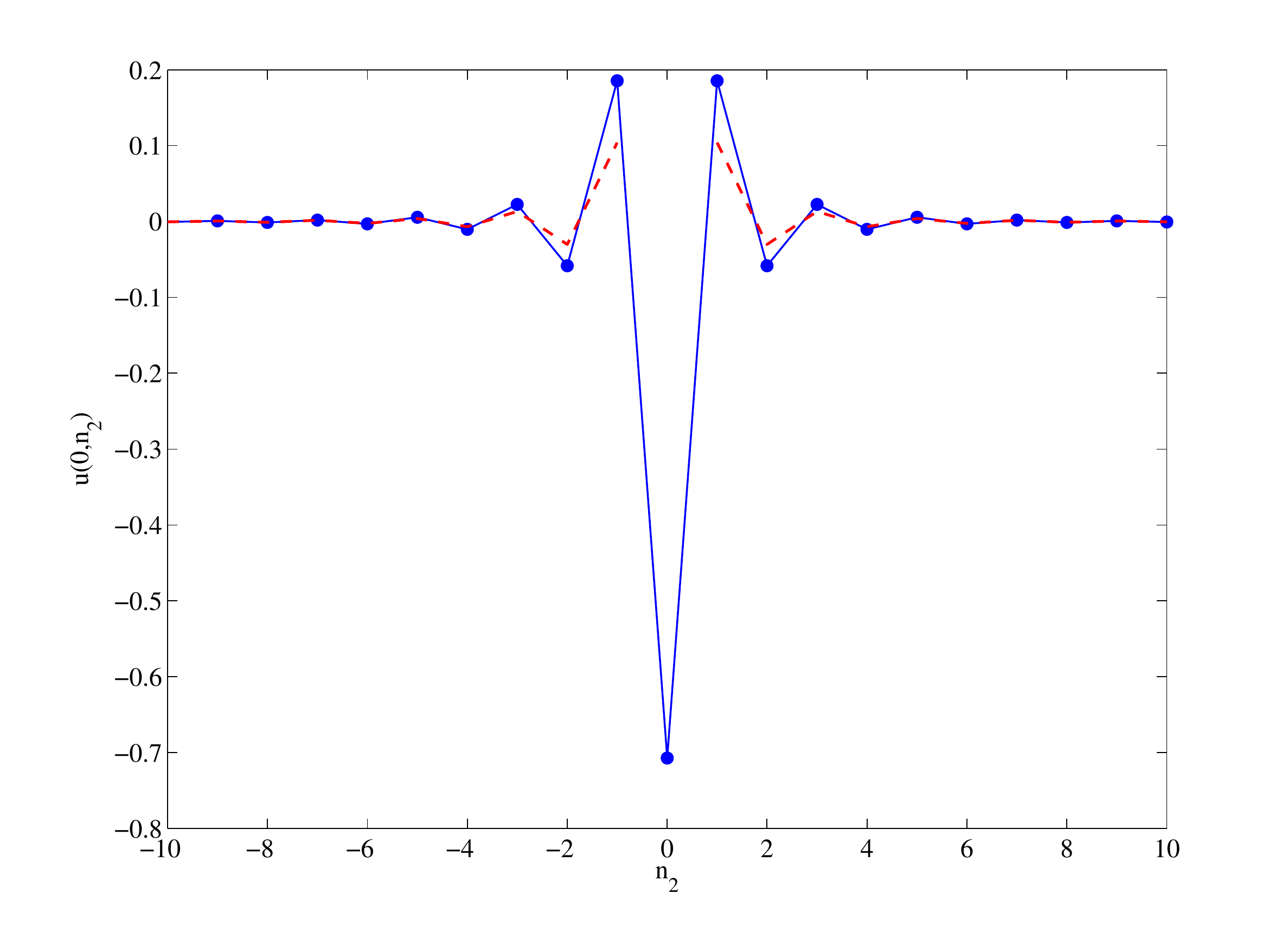}
}\qquad
\subfigure[\label{fig:p1-m2-n1-0}
The field along the line $n_1=0$ for the skew-symmetric mode]{
\includegraphics[width=0.35\linewidth]{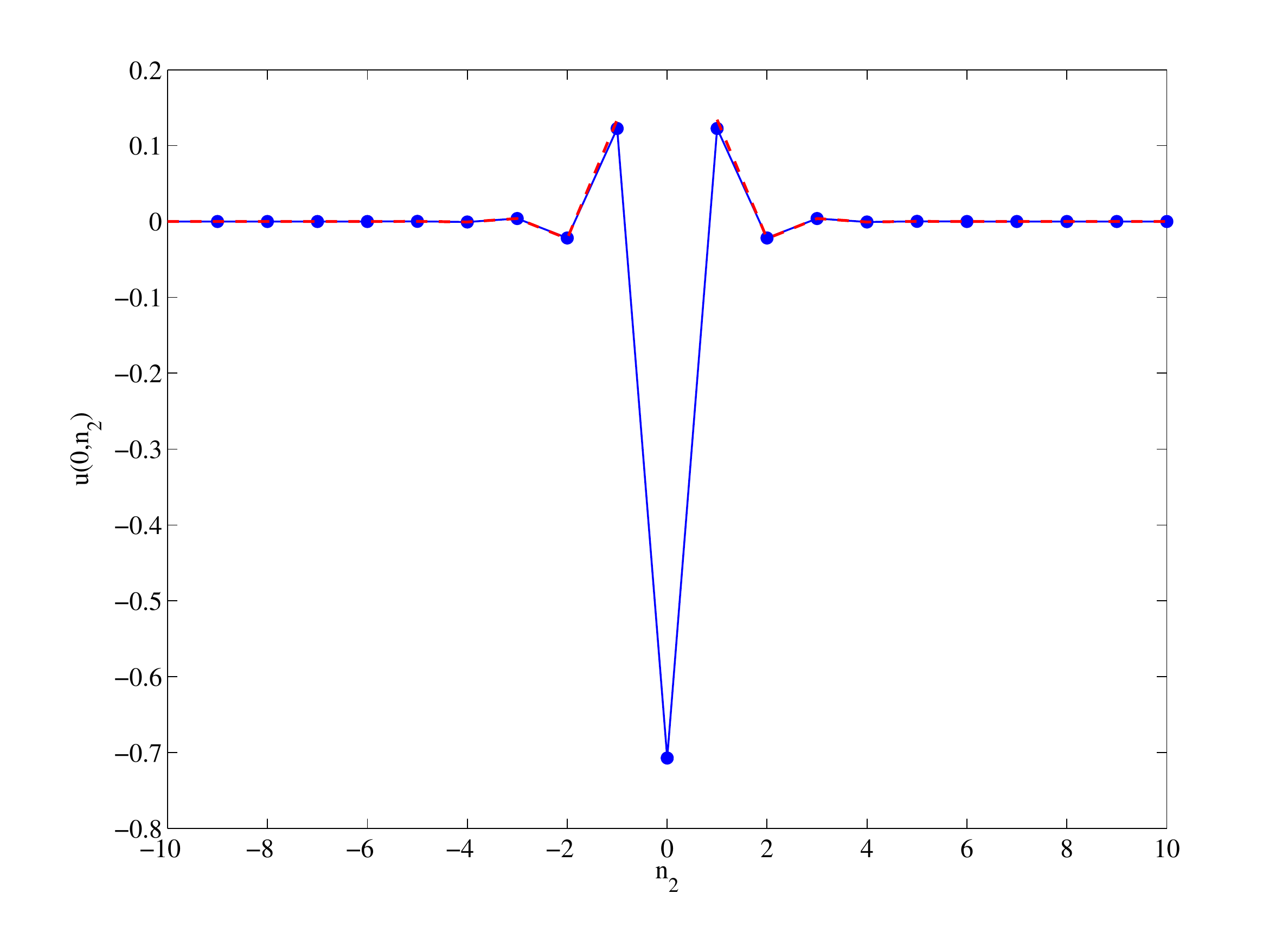}
}\qquad
\caption{\label{fig:p1}
The localised defect mode for a pair of defects with $r=0.49$.
The solid curves are the out-of-plane displacement along the indicated line, and the dashed curves are the associated asymptotic expansions in the far field (cf. equations~\eqref{eq:on-force-line} and~\eqref{eq:perp-force-line} as appropriate).
The dash-dot curve in  figure~\ref{fig:p1-m1-n2-0} shows the band edge expansion (cf. equation~\eqref{eq:band-edge-field-n2-0}).}
\end{figure}

Figure~\ref{fig:p1} shows the two defect modes together with the field along the lines $n_1=0$, and $n_2=0$ and the associated asymptotic field at infinity.
In addition, the dash-dot line in figure~\ref{fig:p1-m1-n2-0} shows the band edge expansion in the vicinity of $\alpha=2$.
In this case, figure~\ref{fig:p1-m1-n2-0} corresponds to value of $\alpha\approx2.025$.
Once again, the asymptotics are in good agreement with the computed field.
Due to the symmetry, the field along the line $n_1=1$ is identical to that in figure~\ref{fig:p1-m1-n1-0} for the symmetric case and identical up to a reflection in the line $u_{0,n_2}=0$ in figure~\ref{fig:p1-m2-n1-0} for the skew-symmetric case.

The lower solid curve in figure~\ref{fig:c-p1} corresponds to $r_{2,1}$ as defined in~\eqref{eq:p1-r1}.
The maximum value of the lower solid curve is given by
\begin{equation}
r^{(\text{max})}_{2,1} = \lim_{\omega\to\sqrt{8}^+} r_{2,1} = \frac{1}{2}.
\end{equation}
Hence for a pair of defects, a symmetric localised mode cannot be initiated for $r\geq1/2$.

\subsection{A triplet of defects}
\label{subsec:triplet}
For the case of three defects, the $3\times3$ matrix $\mathcal{G}(\omega)$ has the $[\mathcal{G}]_{11}$ and $[\mathcal{G}]_{12}$ elements as defined in equations~\eqref{eq:G11} and~\eqref{eq:G12}.
The remaining independent component is
\begin{equation}
\left[\mathcal{G}(w)\right]_{13} =
\left[\mathcal{G}(\omega)\right]_{11} - \frac{\alpha}{2} +\frac{\alpha}{\pi}E\left(\frac{4}{\alpha^2}\right),
\label{eq:G13}
\end{equation}
where $E(x)$ is the complete Elliptic Integral of the second kind.
The solutions of the solvability condition are of similar form to the previous two cases and are omitted for brevity.
The high frequency asymptotics for $r(\omega)$ are
\begin{equation}
r_{3,1}\sim \frac{4-\sqrt{2}}{\omega^2},\qquad r_{3,2}\sim\frac{4}{\omega^2},\qquad\text{and}\qquad
r_{3,3}\sim \frac{4+\sqrt{2}}{\omega^2}\qquad\text{as}\qquad \omega\to\infty,
\end{equation}
which again coincide with the solvability condition for~\eqref{eq:spec-line} for the case of a particle triplet ($N=3$).
The maximum values of $r_{3,i}(\omega)$ are $r_{3,1}^{(\text{max})} = 1-3\pi/16$, $r_{3,2}^{(\text{max})} = 7/8-(8-4\pi)^{-1}$, and $r_{3,3}^{(\text{max})} = 1$.

\label{subsec:triplet-2}

For convenience, the three localised eigenmodes, along with plots of the associated asymptotic expressions are shown in figures~\ref{fig:p2}--\ref{fig:p2-f3} in appendix A.
Plots of the displacement field along the lines $n_2=0$, $n_1=1$ and $n_1=0$ are shown.
The dash-dot line in figure~A.10(b) shows the band edge expansion in the vicinity of $\alpha=2$.
In this case, figure~A.10(b) corresponds to value of $\alpha\approx2.017$.
There are two symmetric modes (the lowest and highest frequency modes) and a single skew-symmetric mode, as expected from the properties of $\mathcal{G}$ discussed in the previous subsection.
However, for defects of mass $r\geq r_{3,1}^{(\text{max})}$, it is not possible to initiate the lower frequency symmetric eigenmode and only a further symmetric mode and a skew-symmetric mode persist.
For values of $r\geq r_{3,2}^{(\text{max})}$, it is only possible to initiate the highest frequency symmetric mode.

\section{An infinite line defect embedded in a uniform lattice}
\label{sec:infinite}

\begin{SCfigure}[1]
\centering
\includegraphics[scale=0.8]{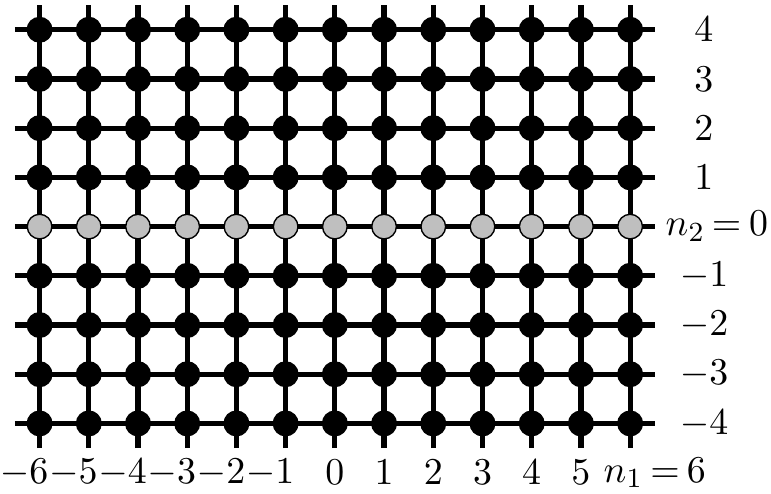}
\caption{A square cell lattice containing an infinite chain of defects with non-dimensional mass $r$ along $n_2=0$, and an ambient lattice composed of particles with unit mass.
As before, the stiffness and length of the links are taken as natural units.}
\label{mass_chain}
\end{SCfigure}

Recently, \cite{Osharovich_etal} studied localised defect modes in square lattices containing infinite defects.
In this section, novel results are presented for homogenisation approximations of long defects.
The dispersion equations for an infinite line of defects in a uniform square lattice are derived and discussed in detail.

The problem of an infinite line of defects embedded in a uniform square lattice, as shown in figure \ref{mass_chain}, is now considered.
%Similar analysis has been carried out by~.
Given the symmetry about the line $n_2=0$, it is convenient to reduce the problem to a half-plane problem, which may be formulated as follows.

\subsection{Equations of motion}\label{sec:eqmotion}
The equation of motion for a particle with $\vec{n}\in\mathbb{Z}\times\mathbb{Z}^+$ is
\begin{subequations}
\begin{equation}
\label{nge1}
\frac{\D^2 u_{\vec{n}}}{\D t^2}= u_{\vec{n}+\vec{e}_1}+ u_{\vec{n}-\vec{e}_1}+ u_{\vec{n}+\vec{e}_2}+ u_{\vec{n}-\vec{e}_2} - 4u_{\vec{n}},
\end{equation}
and for $n_1\in\mathbb{Z}$, $n_2=0$ is
\begin{equation}
\label{ne0}
r\frac{\D^2 u_{n_1,0}}{\D t^2}= u_{n_1+1,0}+ u_{n_1-1,0}+ u_{n_1,1}+ u_{n_1,-1} - 4u_{n_1,0}.
\end{equation}
\end{subequations}
Since the geometry in figure \ref{mass_chain} is periodic in $n_1$, and time-harmonic solutions are of primary interest, the solution $u_{\vec{n}}$ is sought in the form
\begin{equation}\label{form}
u_{\vec{n}}=U_{n_2}e^{\text{i}(n_1\kappa-\omega t)}\;,
\end{equation}
where $\omega>0$  is the angular frequency,  $\kappa\in\mathbb{R}$ is the non-dimensional Bloch parameter (normalised by the length of the lattice bonds) in the $n_1$ direction, and $U_{n_2}$ is the amplitude. 
Assuming the form~\eqref{form},  the equations of motion~\eqref{nge1} and~\eqref{ne0} may be written
\begin{equation}\label{eqF1}
U_{n_2+1}+U_{n_2-1}-2\Omega_1(\kappa, \text{i}\omega)U_{n_2}=0\quad  (n_2\ge 1)\;,
\end{equation}
\begin{equation}\label{eqF2}
U_{1}+U_{-1}-2\Omega_r(\kappa,  \text{i}\omega)U_0=0\;,
\end{equation}
where 
\begin{equation}\label{Oa}
\Omega_\beta(\kappa, z)=1+2\sin^2(\kappa/2)+\frac{\beta z^2}{2}\;.
\end{equation}
\paragraph{The solution in the upper half-plane $(n_2>0)$.}
The solution for $n_2\ge 2$  is then sought in the form
 \begin{equation}\label{eqUF}
 U_{n_2}=\lambda^{n_2}U_1\;, \qquad |\lambda|\le 1\;.
 \end{equation}
The condition $|\lambda| <1$ imposes the localised displacement field about the chain of masses along $n_2=0$.
The case of $|\lambda|=1$ corresponds to  a field which propagates sinusoidally, with constant amplitude,  away from  $n_2=0$ in the transverse direction.
Together, equations~\eqref{eqF1} and~\eqref{eqUF} imply
 \begin{equation}\label{eqlam}
 \lambda^2-2\Omega_1(\kappa,  \text{i}\omega)\lambda+1=0\;. 
 \end{equation}
 The solution of~\eqref{eqlam}  is
 \begin{equation}\label{sollam}
 \lambda=\left\{\begin{array}{ll}
 \Omega_1(\kappa, \text{i}\omega)-\text{sign}(\Omega_1(\kappa, \text{i}\omega))\sqrt{\Omega_1(\kappa, \text{i}\omega)^2-1}&\quad \text{ for }|\Omega_1(\kappa, \text{i}\omega)|>1\;,\\
 \pm 1&\quad \text{ for } \Omega_1(\kappa, \text{i}\omega)=\pm 1\;,\\
 \Omega_1(\kappa, \text{i}\omega)\pm\text{i}\sqrt{1-\Omega_1(\kappa, \text{i}\omega)^2} &\quad \text{ for } |\Omega_1(\kappa, \text{i}\omega)|<1\;.\end{array}\right.
 \end{equation}
It follows from (\ref{sollam}) that if $|\Omega_1(\kappa,  \text{i}\omega)|>1$,  then $|\lambda|<1$.
Further, if $|\Omega_1(\kappa, \text{i}\omega)|\le 1$, then $|\lambda|=1$.

\subsection{Skew-symmetric modes}\label{sec:anti-sym} 
The solution for the problem when the skew-symmetry conditions are imposed along $n_2=0$, is zero. Indeed, for the case of a mode, skew-symmetric about $n_2=0$, the condition $u_{n_1,n_2} = -u_{n_1,-n_2}$ is imposed, and displacements are zero along the defect. Therefore, the solution of (\ref{nge1}) which is zero along $n_2=0$ and either decays or propagates with constant amplitude at infinity, is the trivial solution.
%Hence, $u_{n_1,0} = -u_{n_1,0}$, whence $U_0=0$ and $U_1=-U_{-1}$ and~\eqref{eqF2} is satisfied.

%For the case of $n_2=1$, substitution of~\eqref{eqUF} into~\eqref{eqF1} yields
% \begin{equation}
%\label{eq:skew-symm-u1}
%(\lambda-2\Omega_1(\kappa, \text{i}\omega))U_1=0\;,
%\end{equation}
%where $U_0=0$ has been used.
%From equation~\eqref{eq:skew-symm-u1}, it follows that a nontrivial solution for $U_1$ requires
% \begin{equation}\label{eqdet1}
 %\lambda-2\Omega_1(\kappa, \text{i}\omega)=0\;.
 %\end{equation}
 %Since it is imposed that $|\lambda|\le 1$, equation~\eqref{eqdet1} implies that $|\Omega_1(\kappa, \text{i}\omega)|\le 1/2$ in order for there to be a solution.
%According to \eqref{sollam}, if $|\lambda|<1$ then $|\Omega_1(\kappa, \text{i}\omega)|>1$.
%Hence, the only remaining case to treat is $|\lambda|=1$ and $|\Omega_1(\kappa, \text{i}\omega)|\le 1/2$.
%In this case, equation~\eqref{sollam} implies that $\lambda$ is complex ($|\Omega_1(\kappa, \text{i}\omega)|< 1$).
%However, for $\lambda\in\mathbb{C}$ to be a solution of~\eqref{eqdet1} requires that $\Omega_1(\kappa, \text{i}\omega)$ is also complex with modulus not greater than 1/2.
%For $\kappa,\omega\in\mathbb{R}$, $\min_{\kappa,\omega}\Omega_1=1$ (cf. equation~\eqref{sollam}) and hence, there exist no real solutions of (\ref{eqdet1}) for $\omega$.
%Thus, for the case of skew-symmetry, there exists only the trivial solution.
 
 \subsection{Symmetric mode}\label{sec:sym}
\paragraph{Dispersion relation of the chain at $n_2=0$ for the symmetric mode.}For the case when symmetry conditions are imposed about $n_2=0$, it will be shown that the dispersion relation for defect modes supported by the infinite line defect
%and not by the ambient lattice
is given by 
\begin{equation}
\label{eqdisploc_1}
\begin{split}
\omega^{(-)}(\kappa)& =\left\{\frac{2}{r(2-r)}\left[\phantom{\sqrt{1^2}}\hspace{-1.8em}1+2\sin^2(\kappa/2)\right.\right.\\
& \hspace{5em}
\left.\left.+ \sqrt{1+4(1-r)^2 \sin^2(\kappa/2)(1+\sin^2(\kappa/2))}\right]\right\}^{1/2}\;.
\end{split}
\end{equation}
%As for the skew-symmetric mode, equations~\eqref{eqUF}--\eqref{sollam} hold for $n_2\ge 2$.
This dispersion relation is determined in two parts. First, the  symmetry conditions are imposed about the line $n_2=0$ and a system is derived which links the displacements along the rows $n_2=0$ and $n_2=1$. Then, the solvability of this system is considered for various cases of $\lambda$, and (\ref{eqdisploc_1}) is deduced.
\paragraph{The system for the displacements $U_0$ and $U_1$.}
For symmetric modes, the condition $u_{n_1,n_2}=u_{n_1, -n_2}$ is imposed for $n_2\ge 0$.
In terms of the amplitude field the symmetry condition for $n_2=1$ is  $U_1=U_{-1}$, whence equation~\eqref{eqF1} for $n_2=1$ and equation~\eqref{eqF2} give the system
\begin{equation}\label{equseful}
U_{0}+(\lambda-2\Omega_1(\kappa,  \text{i}\omega))U_1=0\;,
\end{equation}
\[U_{1}-\Omega_r(\kappa,  \text{i}\omega)U_0=0\;,\]
where~\eqref{eqUF} has already been used.
It is convenient to introduce the following matrix notation
\begin{equation}\label{eqdet2}
\boldsymbol{\mathcal{S}}_{r}(\kappa,  \text{i}\omega)\vec{u}=\vec{0}\;,
\end{equation}
with 
\[\boldsymbol{\mathcal{S}}_{r}(\kappa, \text{i}\omega)=\left(\begin{array}{cc}
\lambda-2\Omega_1(\kappa, \text{i}\omega)& 1\\ \\
1 & -\Omega_r(\kappa, \text{i}\omega)\end{array}\right) \quad \text{ and }\quad \vec{u}=[U_1, U_0]^\mathrm{T}\;.\] 
For non-trivial solutions $U_j$ (with $j=0,1$) of (\ref{eqdet2}), it is required that
\[\text{det}(\boldsymbol{\mathcal{S}}_{r}(\kappa, \text{i}\omega))=0\;,\]
which leads to
\begin{equation}
\label{eq:deta}
\frac{\Omega_r(\kappa,  \text{i}\omega)}{\lambda}(\lambda^2-2\Omega_1(\kappa,  \text{i}\omega)\lambda)+1=0\;.
\end{equation}
Together with~\eqref{eqlam}, equation~\eqref{eq:deta} yields
\begin{equation}\label{eqmcdet}
\lambda-\Omega_r(\kappa,  \text{i}\omega)=0\;.
\end{equation}
\paragraph{Solutions of equation (\ref{eqmcdet}).}
In this part, it is shown that for $0<r<1$, there exist no solutions of the symmetric problem when  $|\lambda|=1$, whereas for $|\lambda|<1$, the dispersion relation (\ref{eqdisploc_1}) can be retrieved. %\subsection{Dispersion relations}\label{sec:disprel}
For the solvability of (\ref{eqmcdet}),
 several cases are now discussed in detail:

\vspace{0.1in}{{{\bf\emph{The case of $|\lambda|= 1$}.\rm} } Here it is proved that there exist no solutions of (\ref{eqmcdet}) for $|\lambda|=1$ and  $\omega>0$.

Firstly, consider the case when $\lambda=\pm 1$. According to (\ref{sollam}), this corresponds to $\Omega_1(\kappa, \text{i}\omega)=\pm 1=\lambda$.
Substitution of this into  (\ref{eqmcdet}), leads to 
\[\Omega_1(\kappa,  \text{i}\omega)-\Omega_r(\kappa,  \text{i}\omega)=0\;.\]

It follows that $\Omega_1(\kappa, \text{i}\omega)=\Omega_r(\kappa, \text{i}\omega)$ if and only if $r =1$.
Thus for $0<r <1$, equation~\eqref{eqmcdet} has no solutions for $\omega>0$.
Therefore, there are no solutions with constant amplitude for the case $\lambda=\pm 1$ and $0<r<1$.
The case of $r=1$ corresponds to an intact lattice and one would expect constant amplitude (Bloch wave) solutions to be supported.

%\paragraph{Solution with constant amplitude: \emph{The case of $\lambda\in\mathbb{C}\setminus\mathbb{R}$}. }
%It remains to deal with
The case of complex $\lambda$ remains. For $\lambda$ to be complex, the condition $|\Omega_1(\kappa, \text{i}\omega)|<1$ has to be satisfied, and then $|\lambda|=1$. Since $r\ne 1$, and $\Omega_r(\kappa, \text{i}\omega)$ is real for $\omega> 0$ and $\kappa\in\mathbb{R}$, then (\ref{eqmcdet}) has no real solutions for $\omega$. 

%Note that $|\Omega_1(\kappa, \text{i}\omega)|<1$ is satisfied for 
%\[\omega^{(1)}(\kappa)<\omega<\omega^{(2)}(\kappa)\;.\]

%\paragraph{Localised defect modes: \emph{The case of $|\lambda|<1$}.}
\vspace{0.1in}{{\bf\emph{The case of $|\lambda|< 1$}. \rm}}
%An extended equation of (\ref{eqmcdet}) is now obtained that will lead to the dispersion relation (\ref{eqdisploc_1}).
Together, equations~\eqref{eqlam} and \eqref{eqmcdet} yield
\[\Omega_r(\kappa,  \text{i}\omega)^2-2\Omega_r(\kappa,  \text{i}\omega)\Omega_1(\kappa,  \text{i}\omega)+1=0\;,\]
which by  (\ref{Oa}), is equivalent to
\begin{equation}\label{biquadom}
\frac{1}{4}r(r-2)\omega^4+(1+2\sin^2(\kappa/2))\omega^2-4\sin^2(\kappa/2)(1+\sin^2(\kappa/2))=0\;.
\end{equation}
Equation~\eqref{biquadom} is a biquadratic equation in terms of $\omega$.
Now, according to~\eqref{sollam}, $|\lambda|<1$ occurs when $|\Omega_1(\kappa,  \text{i}\omega)|>1$.
Furthermore, for $|\lambda|<1$ equation~\eqref{eqmcdet} implies that $|\Omega_r(\kappa,  \text{i}\omega)|<1$.
Moreover, the inequalities $|\Omega_r(\kappa,  \text{i}\omega)|<1$ and  $|\Omega_1(\kappa,  \text{i}\omega)|>1$,  lead to
\begin{equation}
\label{ineq1}
r^{-1/2}\omega^{(1)}(\kappa)<\omega<r^{-1/2}\omega^{(2)}(\kappa)
\end{equation}
together with
\begin{equation}\label{ineq2}
\omega^{(1)}(\kappa)> \omega \quad  \text{ or }\quad \omega >\omega^{(2)}(\kappa)
\;,
\end{equation}
where 
\begin{subequations}
\begin{equation*}\label{Om1ma}
\omega^{(1)}(\kappa)=2|\sin(\kappa/2)|\;,
%\quad(\Omega_1(\kappa,  \text{i}\omega)=1)\;,
%\end{equation*}
%\begin{equation*}\label{Om2ma}
\quad \text{ and }\quad\omega^{(2)}(\kappa)=2\sqrt{1+\sin^2(\kappa/2)}\;.
%\quad(\Omega_1(\kappa,  \text{i}\omega)=-1)\;.
\end{equation*}
\end{subequations}
Either of the 
%last two 
inequalities (\ref{ineq2}) implies that $|\lambda|<1$, and when one of these inequalities is taken with \eqref{ineq1}, the solutions of \eqref{eqmcdet} for $\omega$ should satisfy these conditions.
Equation~\eqref{biquadom} should be solved subject to conditions~\eqref{ineq1} and~\eqref{ineq2} in order to determine the dispersion equations.

%whence the following solutions are found
%\begin{subequations}
%\begin{equation}\label{Om1ma}
%\omega^{(1)}(\kappa)=2|\sin(\kappa/2)|\;,\quad(\Omega_1(\kappa,  \text{i}\omega)=1)\;,
%\end{equation}
%or
%\begin{equation}\label{Om2ma}
%\omega^{(2)}(\kappa)=2\sqrt{1+\sin^2(\kappa/2)}\;, \quad(\Omega_1(\kappa,  \text{i}\omega)=-1)\;.
%\end{equation}
%\end{subequations}
%Further, substitution of $\lambda=\pm 1$ into (\ref{eqmcdet}), for $r\neq1$, 
%gives %equation~\eqref{eqmcdet} has 
%the following solutions
%\begin{subequations}
%\begin{equation}
 %\omega^{(3)}(\kappa)=r^{-1/2}\omega^{(1)}(\kappa)\;,\qquad (\Omega_{r}(\kappa, \text{i}\omega)=1)\;,
%\end{equation}
%or
%\begin{equation}
%\omega^{(4)}(\kappa)=r^{-1/2}\omega^{(2)}(\kappa)\;,\qquad (\Omega_{r}(\kappa, \text{i}\omega)=-1).
%\end{equation}
%\end{subequations}

Firstly, note that for localised modes, inequalities~\eqref{ineq1} and~\eqref{ineq2} yield either 
\[\omega^{(2)}(\kappa)< r^{-1/2}\omega^{(2)}(\kappa)\;,\quad  \text{ for all }\kappa\in\mathbb{R}\;,\]
 which leads to $r<1$, or 
 \[ r^{-1/2}\omega^{(1)}(\kappa)<\omega^{(1)}(\kappa)\;, \quad \text{ for all }\kappa\in\mathbb{R}\;,\]
 which is never satisfied for any $0<r<1$.

\vspace{0.1in}{\bf \emph{Roots of the biquadratic equation~\eqref{biquadom}.}\rm}
The solutions of ~\eqref{biquadom} are as follows:
\begin{equation}\label{eqdisploc_1a}
\begin{split}
\omega^{(\pm)}(\kappa) & =\left\{\frac{2}{r(2-r)}\left[1+2\sin^2(\kappa/2)\right.\right.\\
& \hspace{5em}
\left.\left.\mp \sqrt{1+4(r-1)^2 \sin^2(\kappa/2)(1+\sin^2(\kappa/2))}\right]\right\}^{1/2}\;,
\end{split}
\end{equation}
for $0<r<1$. Here, it is shown that $\omega^{(+)}$ is not a solution of (\ref{eqmcdet}), whereas $\omega^{(-)} $  is a solution of this equation.
%(\ref{eqdisploc_1}).

%and
%\begin{equation}\label{eqdisploc_2a}
%\omega^{(+)}(\kappa)= 2|\sin(\kappa/2)|\sqrt{\frac{ 1+\sin^2(\kappa/2)}{1+2\sin^2(\kappa/2)}}\;,
%\end{equation}
%for $r=0$.
%Using (\ref{eqdisploc_1a}), for $r\to 0$:
%\[(\omega^{(+)}(\kappa))^2=\frac{4\sin^2(\kappa/2)(1+\sin^2(\kappa/2))}{1+2\sin^2(\kappa/2)}+O\left(r\right)\;.\]
%Here the leading order term is equivalent to~\eqref{eqdisploc_2a}.

%\paragraph{Roots of the solvability condition: Equation~\eqref{eqmcdet}.}
%Now it will be checked which of the  solutions~\eqref{eqdisploc_1a} of the biquadratic equation~\eqref{biquadom}, satisfy the inequalities~\eqref{ineq1} and either~$\eqref{ineq2}_1$ or ~$\eqref{ineq2}_2$ and, consequently, are roots of equation~\eqref{eqmcdet}. 

\vspace{0.1in}{{\bf\emph{The function $\omega^{(-)}$. }\rm} Here it is proved that  $\omega^{(-)}$ satisfies $(\ref{ineq1})$ and $\eqref{ineq2}_2$, and is therefore a root of~\eqref{eqmcdet}. 

Using Young's inequality,
%$ab\le 2^{-1}(a^2+b^2)$, 
\[\sqrt{1+4(r-1)^2\sin^2(\kappa/2)(1+\sin^2(\kappa/2))}\le 1+2(r-1)^2\sin^2(\kappa/2)(1+\sin^2(\kappa/2))\;,\]
and so from (\ref{eqdisploc_1a})
\begin{eqnarray*}
(\omega^{(-)}(\kappa))^2&\le& \frac{4}{r(2-r)}(1+\sin^2(\kappa/2))(1+(r-1)^2\sin^2(\kappa/2))\\
&\le &\frac{4}{r(2-r)}(1+\sin^2(\kappa/2))(1+(r-1)^2)\;,
\end{eqnarray*}
for any $\kappa\in \mathbb{R}$.
The last factor on the right hand side is positive and convex, and for $0<r<1$ is less than $2-r$. Therefore 
\begin{equation}\label{S1}
\omega^{(-)}(\kappa)< r^{-1/2}\omega^{(2)}(\kappa)\;,\quad  \text{ for all }\kappa\in \mathbb{R}\;.
\end{equation}
Now, the function inside the radical of $(\omega^{(-)})^2$ (see (\ref{eqdisploc_1a})), can be written as 
\[(r-1)^2(2\sin^2(\kappa/2)+1)^2+1-(r-1)^2\;,\]
and this implies that for $0<r<1$ and $\kappa\in \mathbb{R}$,
\[(\omega^{(-)}(\kappa))^2>\frac{2}{r}(1+2\sin^2(\kappa/2))\;.\]
This then leads to 
\begin{equation}\label{S2}
\omega^{(-)}(\kappa)> r^{-1/2}\omega^{(1)}(\kappa)\;, \quad  \kappa\in \mathbb{R}.
\end{equation}
%Also,
It remains to  show $\omega^{(-)}$ satisfies  $\eqref{ineq2}_2$. Due to the equality $1-(r-1)^2=r(2-r)$, it can be seen that this function is concave and less than 1 for $r<1$.
Then using this fact, $\omega^{(-)}$ can also be estimated from below:
\begin{equation}\label{S3}
\omega^{(-)}(\kappa)>\omega^{(2)}(\kappa)\;, \quad \text{ for all }\kappa \in \mathbb{R}\;.
\end{equation}
Then (\ref{S1}), (\ref{S2}) and (\ref{S3}) show that $\omega^{(-)}$ satisfies inequalities (\ref{ineq1}) and $(\ref{ineq2})_2$ and is therefore a solution of (\ref{eqmcdet}).

\vspace{0.1in}{{\bf \emph{ The function $\omega^{(+)}$.} \rm}
Now it is shown that $\omega^{(+)}$ does not satisfy $\eqref{ineq1}$ and is consequently not a root of~\eqref{eqmcdet}. 
%Now consider he function $\omega^{(+)}$.
%It will be shown that $\omega^{(+)}$  is not a solution the original equation~\eqref{eqmcdet}, since $|\Omega_r(\kappa, \text{i}\omega^{(+)})|>1$.
Indeed, since the function in the radical of $(\omega^{(+)})^2$ is always positive
\[(\omega^{(+)}(\kappa))^2<\frac{4}{r(2-r)}\sin^2(\kappa/2)\;.\]
{for } $\kappa\in \mathbb{R}$. Then, since $r<1$, it can be asserted that
\[\omega^{(+)}(\kappa)<r^{-1/2}\omega^{(1)}(\kappa)\;, \quad \text{ for } \kappa\in \mathbb{R}\;. \]
Therefore, $\omega^{(+)}$ does not satisfy~\eqref{ineq1} and is not a solution of~\eqref{eqmcdet}.

\paragraph{Analysis of the dispersion relation when $|\lambda|<1$ for various $r$.}
If $\omega$ satisfies 
%The dispersion relation that satisfies
$|\Omega_1(\kappa, \text{i}\omega)|>1$ and $|\Omega_r(\kappa, \text{i}\omega)|<1$ for all $\kappa\in\mathbb{R}$ and is a solution of~\eqref{eqmcdet} for $0<r<1$
%is therefore given by
then the dispersion relation for waves in the infinite wave guide is given by (\ref{eqdisploc_1}).
%\begin{equation}
%\label{eqdisploc_1}
%\begin{split}
%\omega^{(-)}(\kappa)& =\left\{\frac{2}{r(2-r)}\left[\phantom{\sqrt{1^2}}\hspace{-1.8em}1+2\sin^2(\kappa/2)\right.\right.\\
%& \hspace{5em}
%\left.\left.+ \sqrt{1+4(1-r)^2 \sin^2(\kappa/2)(1+\sin^2(\kappa/2))}\right]\right\}^{1/2}\;,
%\end{split}
%\end{equation}
%for $0<r<1$.

For $r\to0$,
\[(\omega^{(-)}(\kappa))^2=\frac{2}{r} (1+2\sin^2(\kappa/2))+\frac{1}{1+2\sin^2(\kappa/2)}+O\left(r\right),\]
where the second term on the right-hand side is bounded and the first term dominates for $r\to 0$.
Thus, $\omega^{(-)}(\kappa)\to\infty$ as $r\to0$.
In figure~\ref{dispmultbeta}, the dispersion relation (\ref{eqdisploc_1}) is plotted for several values of $r$.
The in-phase standing wave solution, of the form~\eqref{form}, is always given when $\kappa=0$ and corresponds to the minima of the dispersion curves.
The frequency of the in-phase standing wave is
\begin{equation}\label{eqinphase}
\omega=\sqrt{\frac{4}{r(2-r)}}\;,
\end{equation}
whereas for the out-of-phase solution, at $\kappa=\pi$ corresponding to the maxima of the dispersion curves, is 
\begin{equation}
\label{eqoutofphase}
\omega=\sqrt{\frac{2}{r(2-r)}\left[3+\sqrt{1+8(1-r)^2}\right]}\;.
\end{equation}

\begin{SCfigure}[1]
\centering
\includegraphics[width=0.5\textwidth]{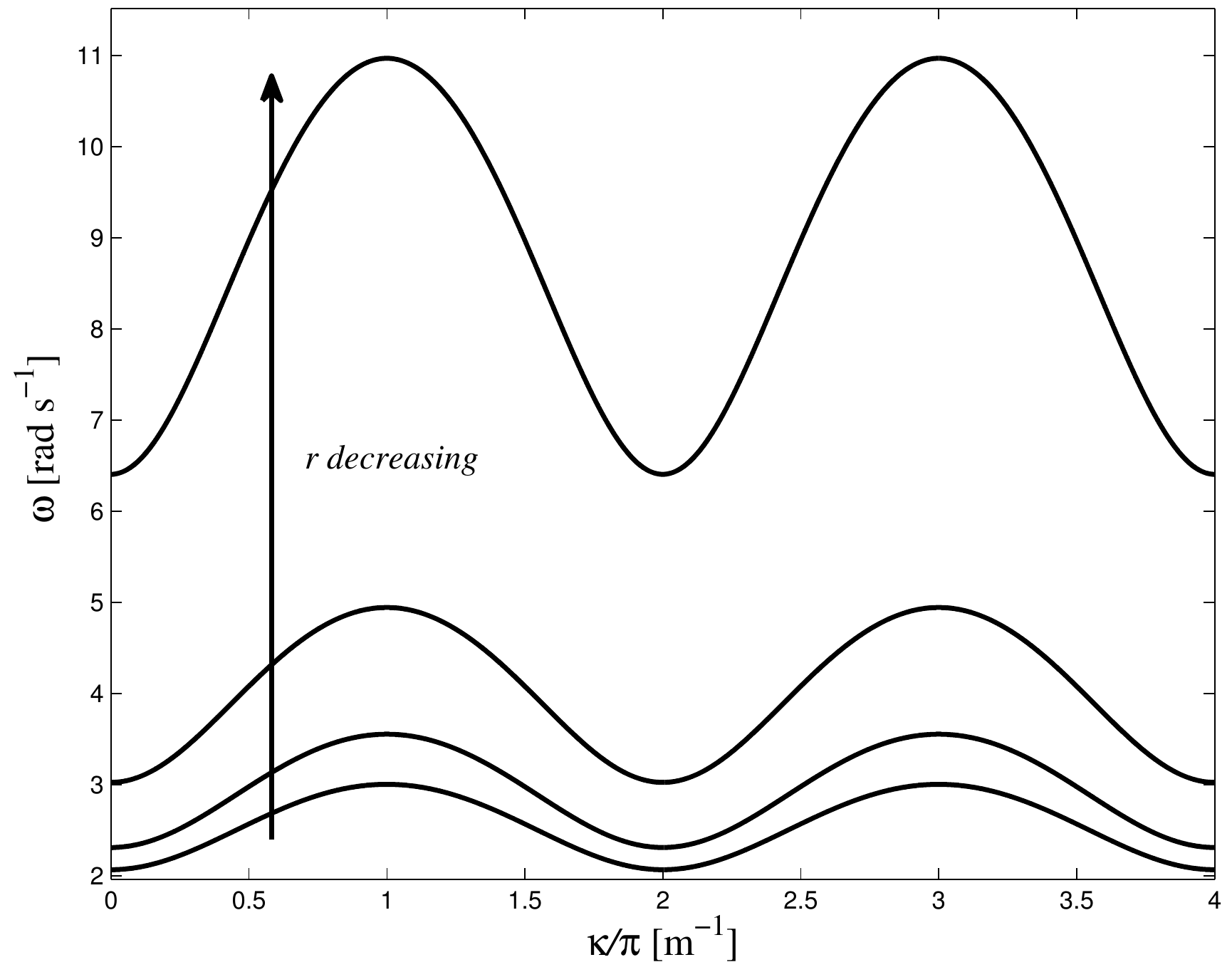}
\centering
\caption{The quantity  $\omega^{(-)}$, given in equation~\eqref{eqdisploc_1}, 
 plotted as a function of the normalised Bloch parameter $\kappa/\pi$ for $r=0.05, 0.25, 0.5$ and $0.75$.}
  \label{dispmultbeta}
\end{SCfigure}

\section{From an infinite inclusion to a large finite defect: The case of large $N$}\label{sec:numsim}

In this section, the objective is to show that the range of eigenfrequencies for which localised eigenmodes exist for the model described in section~\ref{sec:finite}, can be predicted using the model of an infinite chain of defects considered in section~\ref{sec:infinite}.
A defect composed of $N=20$ particles of non-dimensional mass $r=0.25$ is embedded within an infinite square lattice.
The eigenfrequencies of the finite defect were computed using the method described in section~\ref{sec:finite} and are shown as dash-dot, and dashed, lines in figure~\ref{sim}.
In this figure, the eigenfrequency $\omega_\text{min}=3.0374$ corresponds to an in-phase standing wave solution, whereas the frequency $\omega_\text{max}=4.9344$ represents the out-of-phase solution.
The maximum and minimum eigenfrequencies are indicated by the dashed lines in figure~\ref{sim}.

Since $N$ is large, it is useful to consider the model  of an infinite chain embedded in a square lattice. Expressions~\eqref{eqinphase} and~\eqref{eqoutofphase} predict the values of the frequency $\omega$ for which there exist such solutions. 
For the numerical values above, the in-phase solution occurs when $\kappa=0$ and $\omega=3.0237$ and
the out-of-phase solution occurs when $\kappa=\pi$ and $\omega=4.9432$.
These values of the frequency are close to those encountered in the problem of the finite defect for $N=20$.
Moreover, all the eigenfrequencies computed for the finite defect lie within the passband for the infinite defect, as shown in figure~\ref{sim}.

\begin{SCfigure}[1]
\centering
\includegraphics[width=0.5\textwidth]{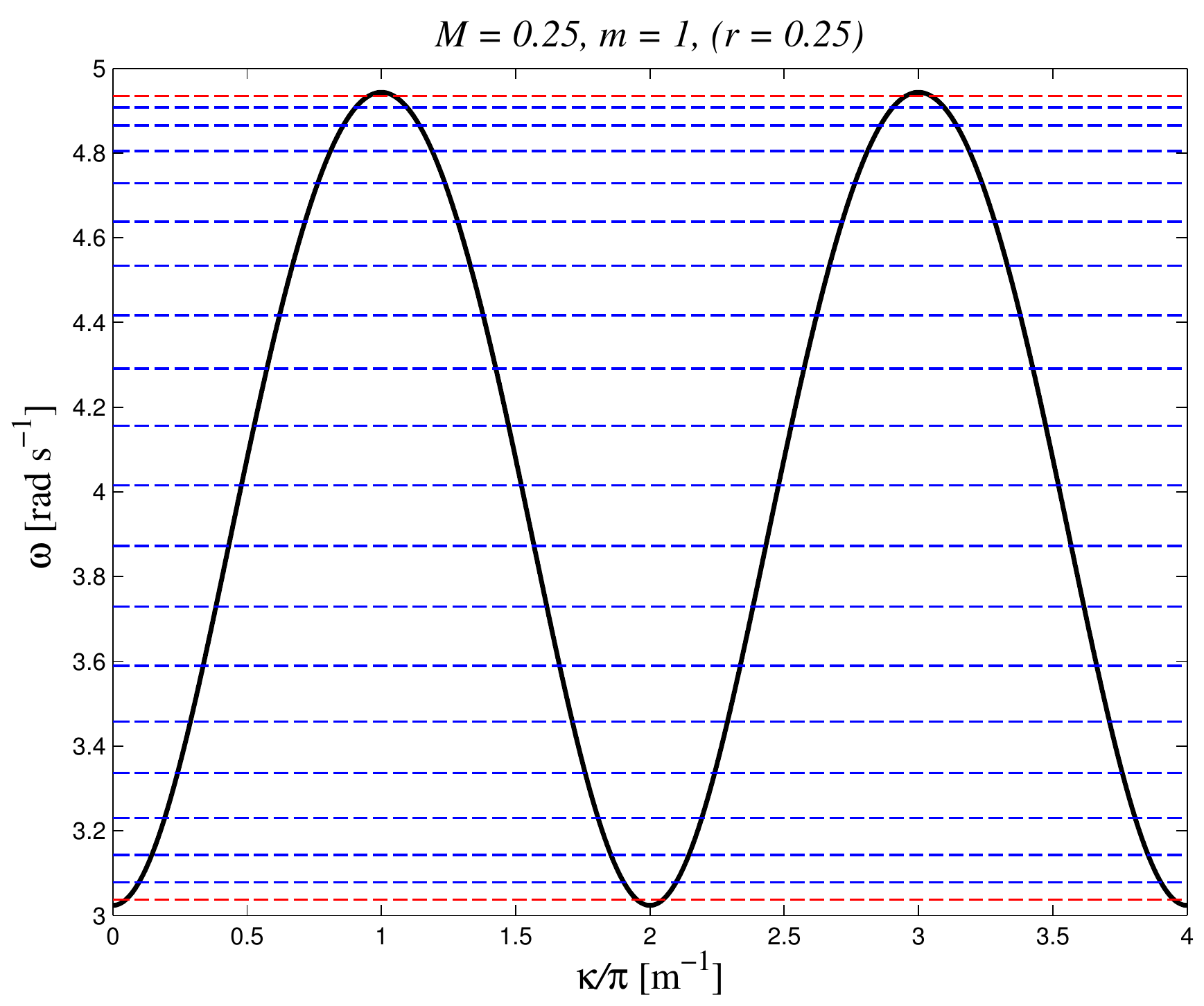}
\centering
\caption{The dispersion equation \eqref{eqdisploc_1}, for the infinite chain, plotted as a function of the normalised Bloch parameter, for $r=0.25$, represented by the solid curve.
Also shown are the dash-dot lines corresponding to the the eigenfrequencies computed for a finite defect containing $N=20$ masses.
The dashed lines correspond to $\omega_\text{min}$ and $\omega_\text{max}$.}
  \label{sim}
\end{SCfigure}

\begin{figure}[htb]
\centering
\subfigure[\label{inphase}
In-phase mode for  $\omega=3.037$]{
\includegraphics[width=0.4\linewidth]{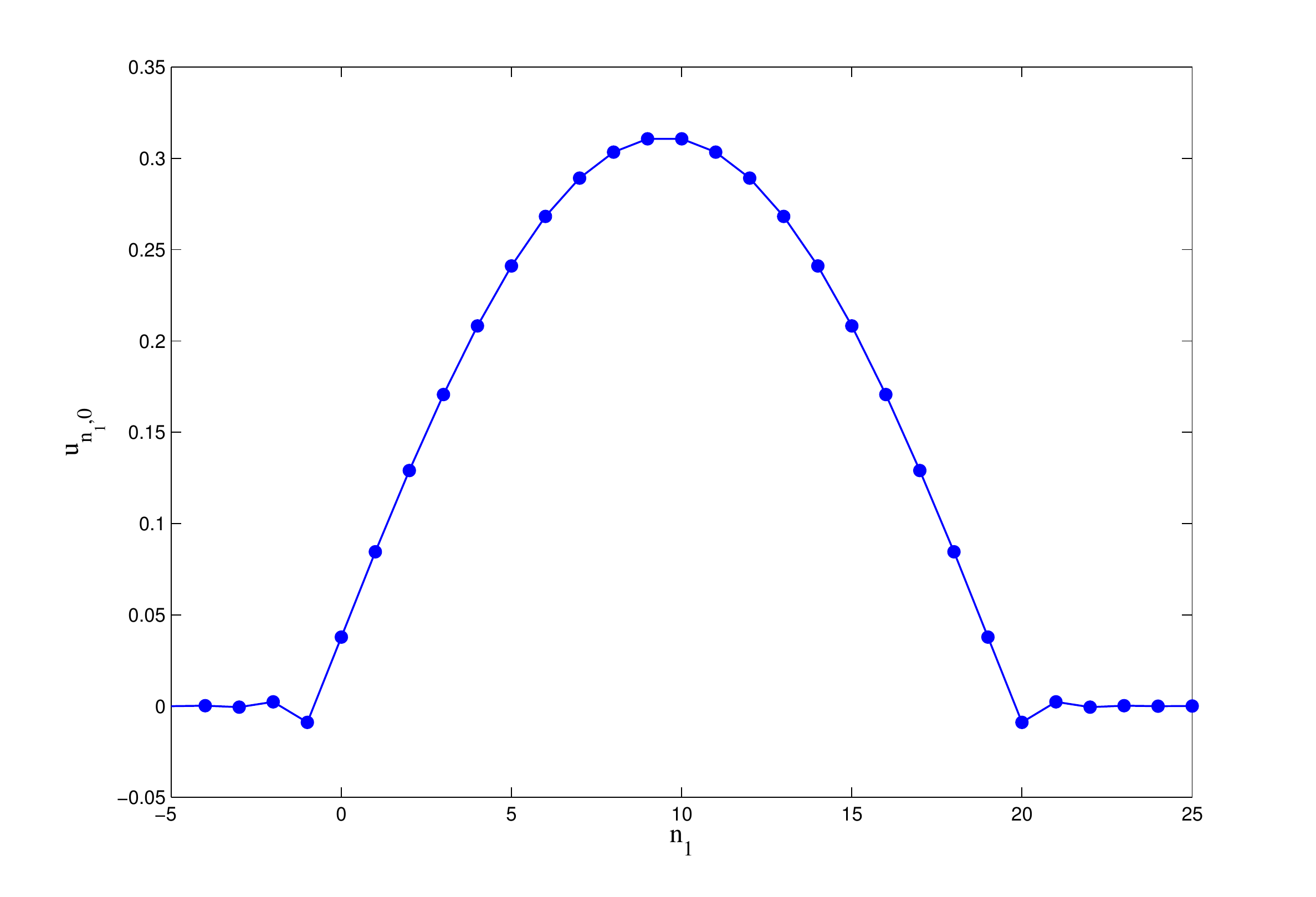}
}\qquad
\subfigure[\label{outofphase}
Out-of-phase mode for $\omega=4.934$]{
\includegraphics[width=0.4\linewidth]{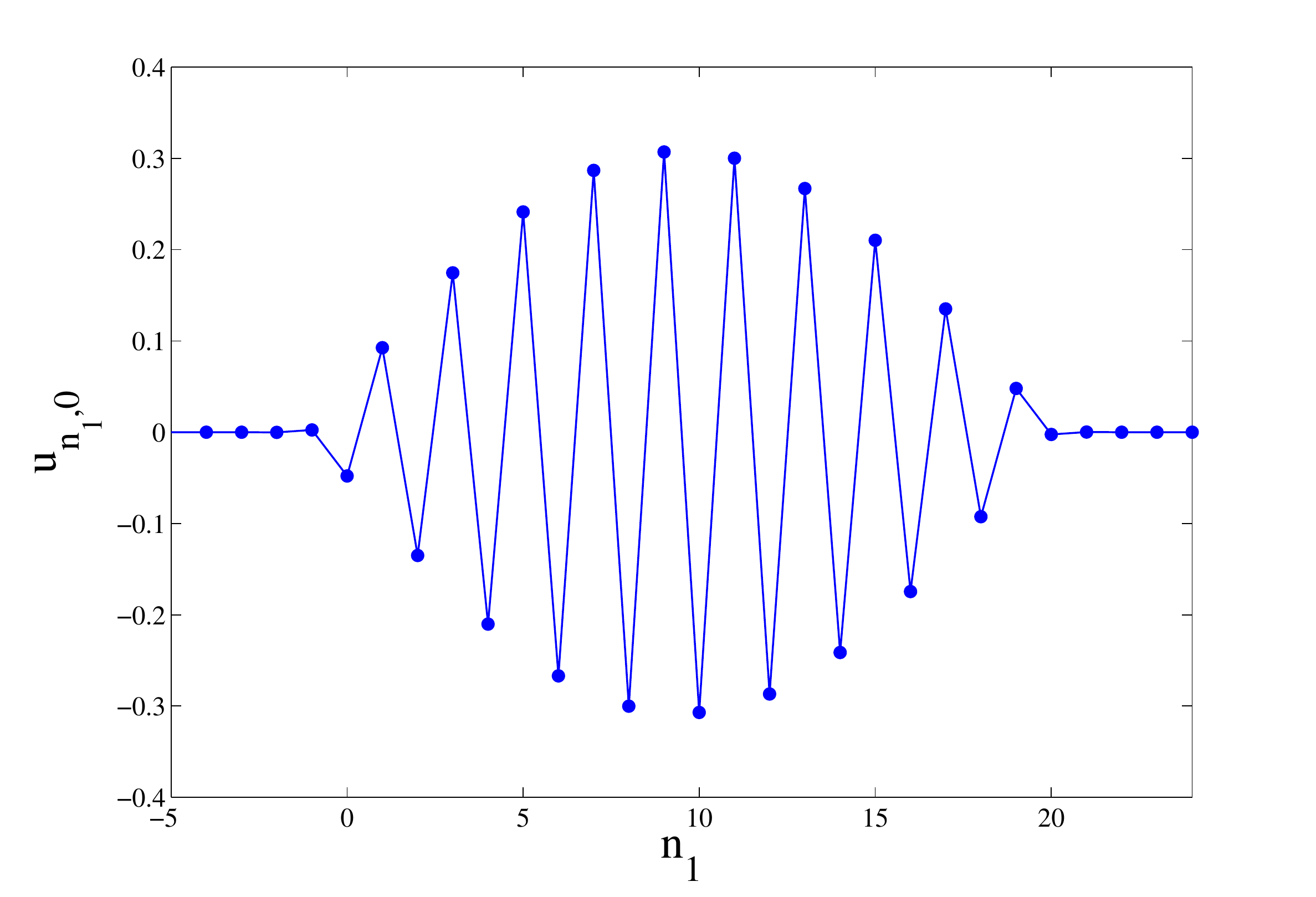}
} 
\caption{The solid lines are the eigenmodes for the maximum and minimum eigenfrequencies for a finite line containing 20 defects.
The envelope function defined in~\eqref{eq:string1} is plotted as the dashed lines.}
 \label{defect_modes}
\end{figure}

Figure \ref{defect_modes} shows the plot of the eigenmodes for the maximum and minimum eigenfrequencies computed for the line defect containing 20 masses. The maximum eigenfrequency $\omega_\text{max}$ corresponds to the out-of-phase mode, whereas the minimum eigenfrequency $\omega_\text{min}$ gives the in-phase mode.

It is remarked that both the field in figure~\ref{inphase}, and the envelope of the field in figure~\ref{outofphase} resemble the first eigenmode of an homogenised rectilinear inclusion. Using this motivation the difference operator
\begin{equation}
\mathcal{D}_{\vec{p}}\left(\cdot\right)_{\vec{p}}
= \left(\cdot\right)_{\vec{p}+\vec{e}_1} + \left(\cdot\right)_{\vec{p}-\vec{e}_1}
+\left(\cdot\right)_{\vec{p}+\vec{e}_2} + \left(\cdot\right)_{\vec{p}-\vec{e}_2}
-4\left(\cdot\right)_{\vec{p}},
\end{equation}
is introduced.
Making use of~\eqref{eq:field}, it is found that
\begin{equation}
\left(\frac{\mathcal{D}_{n_1,0}}{\omega^{2}} +1\right)u_{n_1,0}
= (1-r)\sum_{p=0}^{N-1}u_{p,0}\left(\mathcal{D}_{n_1,0}+\omega^2\right)g(n_1,0,p;\omega),
\end{equation}
where $\vec{n}$ has been restricted to $\{\vec{n}:0\leq n_1 \leq N-1,\,n_2=0\}$.
Since the lattice Green's matrix is a difference kernel (i.e. depends on the difference $|n_{1}-p|$),
\begin{equation}
\left(\frac{\mathcal{D}_{n_1,0}}{\omega^{2}} +1\right)u_{n_1,0}
= (1-r)\sum_{p=0}^{N-1}u_{p,0}\left(\mathcal{D}_{p,0}+\omega^2\right)g(n_1,0,p;\omega),
\end{equation}
whence, and recalling from~\eqref{eq:gov} that $(\mathcal{D}_{\vec{n}}+\omega^2)g(\vec{n},p,\omega) = \delta_{n_1,p}\delta_{n_2,0}$, it is found that
\begin{equation}
(\mathcal{D}_{\vec{n}}+r\omega^2)u_{\vec{n}}=0,\quad\text{for}\quad\vec{n}\in\{\vec{n}:0\leq n_1 \leq N-1,\,n_2=0\}.
\label{eq:fde-on-line}
\end{equation}
It is observed that for a sufficiently large inclusion, the field above and below the inclusion behaves as $u_{n_1,1}=u_{n_1,-1}\approx\lambda u_{n_1,0}$, with $|\lambda|<1$, in a similar manner to the infinite inclusion.
Hence, using~\eqref{eq:fde-on-line} together with the aforementioned approximation yields
\begin{equation}
u_{n_1+1,0} + u_{n_1-1,0}  - 2u_{n,0} +\left[r\omega^2-2\left(1-\lambda\right)\right]u_{n_1,0} \approx 0,
\label{eq:fde2}
\end{equation}
for $0\leq n_1 \leq N-1$.
The first three terms on the left hand side of~\eqref{eq:fde2} correspond to the second order central difference operator.
Hence, introducing the continuous variable $\eta=n_1$ (where the reader is reminded that the length of the lattice links has been normalised to unity) equation~\eqref{eq:fde2} is written as
\begin{equation}
\left[\frac{d^2}{d\eta^2}+r\omega^2-2\left(1-\lambda\right)\right]u(\eta) \approx 0.
\label{eq:de-string}
\end{equation}
The form of equation~\eqref{eq:de-string} suggests that the homogenised system is analogous to a string on an elastic foundation, with the constant $2\left(1-\lambda\right)$ characterising the effective stiffness of the foundation.
It is emphasised that $|\lambda|<1$ and as such, the stiffness of the elastic foundation is positive.

Consider the problem of an infinite inclusion.
According to equations~\eqref{Oa},~\eqref{sollam}, and~\eqref{eqmcdet}, the value of $\lambda$ corresponding to the lowest eigenmode is $\lambda = 1+r\omega^2/2$.
For this value of $\lambda$, the second order derivative vanishes according to equation~\eqref{eq:de-string}.
Moreover, for the displacement at infinity to be finite, $u(\eta)$ must be constant for all $\eta$ and the exact solution for the infinite inclusion is obtained.

For the finite inclusion, it is observed that the displacements at the endpoints are small (cf. figure~\ref{inphase}).
Hence, for a simple estimate it suffices to impose $u(0)=u(N-1)=0$ whence the solution to~\eqref{eq:de-string} is
\begin{equation}
u(\eta) = u_0\sin\left(\sqrt{r\omega^2-2(1-\lambda)}\eta\right),\;\text{with}\;
\lambda = 1+\frac{1}{2}\left[\left(\frac{q\pi}{N-1}\right)^2-r\omega^2\right],
\label{eq:string1}
\end{equation}
where $q$ is an odd number and $u_0$ an arbitrary scaling constant.
The first eigenmode corresponds to $\lambda = -0.1396$, which is close to the mean value of $\lambda$ obtained from the full numerical computation ($\lambda=-0.1426$).
The approximation~\eqref{eq:string1} for $\lambda = -0.1396$ is plotted in figure~\ref{inphase} as the dashed line.
The same approximation is used to produce the envelope function plotted as the dashed lines in figure~\ref{outofphase}.
One may observe that this, relatively simple, homogenised model predicts the envelope of the field very well.

\section{Concluding remarks}

A comparative analysis of two classes of problems has been presented: localised vibrations around a finite size defect created by a line of masses in a square lattice and an infinite waveguide represented by a chain of masses embedded in an ambient lattice.

Although the physical configurations and the methods of analysis of these problems are different, one may observe remarkable properties of solutions, which can be used to make a strong connection.
As illustrated in figure~\ref{sim}, the pass band for frequencies of waveguide modes, localised around an infinite chain of masses in a square lattice, contains all eigenmodes describing vibrations localised around a rectilinear defect built of a finite number of masses embedded into the lattice.

Special attention is given to the band edges: figure~\ref{sim} shows that the frequencies of the eigenmodes for a finite rectilinear defect are distributed non-uniformly and they cluster around the edges of the pass band identified for the infinite waveguide problem.
Furthermore, the limit, as one approaches the band edge frequency, corresponds to a homogenisation approximation of the rectilinear defect as an inclusion embedded into a homogenised ambient system.
The illustrative numerical simulation is produced for an array of 20 masses.
However, the effect shown is generic, and, with an increased number of masses, the density of frequencies of localised modes near the band edges, identified for an infinite waveguide, increases. 

Symmetric and skew-symmetric modes have been constructed and analysed for a rectilinear ``inclusion'' built of a finite number of masses embedded into the lattice.
It has also been shown that the total  force produced by the vibrating discrete inclusion on the ambient lattice is zero for all skew-symmetric modes.
Consequently, the displacement fields, associated with skew-symmetric modes, decay at infinity like dipoles, vanishing faster than the displacements corresponding to symmetric modes.
This follows from the analytical representations for the solutions and illustrated in figures~\ref{fig:p1} and~\ref{fig:p2}  where the skew-symmetric modes  appear to be localised to a much higher degree than symmetric modes.
In the aforementioned numerical simulations, the skew-symmetric and symmetric modes appear in pairs, and the frequency of the skew-symmetric mode is higher than the frequency of the corresponding symmetric mode.
With reference to figure~\ref{fig:phase-diagrams} it is also observed that, in contrast to 1D and 3D cases, a defect mode can be initiated  for any value of the contrast parameter $r\in(0,1)$.
In other words, removing any amount of mass for a point in a square lattice will yield a localised eigenmode.

Finally, the reader's attention is drawn to the symmetric and skew-symmetric eigenmodes for a chain of 20 masses shown in figure~\ref{defect_modes}.
The corresponding frequencies are the maximum and minimum values among the array of frequencies associated with horizontal lines in figure~\ref{sim}.
The envelope curves for both diagrams in figure~\ref{defect_modes} represent the first eigenmode of a  homogenised rectilinear inclusion.
The simple homogenised model presented in section~\ref{sec:numsim} provides the envelope curves for the finite inclusion.
The form of the homogenised system suggests that, macroscopically, the inclusion behaves as a string on an elastic foundation.
As expected, the skew-symmetric mode of figure~\ref{outofphase} has the higher frequency than the symmetric mode of figure~\ref{inphase}. 

\subsection*{Acknowledgements}
{\footnotesize
The authors wish to thank Professor R.C. McPhedran for his invaluable discussions and electronic correspondence.
D.J.C. gratefully acknowledges the financial support of EPSRC through a research scholarship (grant number EP/H018514/1).
M.J.N. acknowledges the financial support of EPSRC (grant number EP/H018239/1).
A.B.M. and N.V.M. gratefully acknowledge the support from the European Union Seventh Framework Programme under contract number PIAP-GA-2011-284544-PARM-2.
}

\appendix
\label{app:triplet-figures}
\section*{Appendix A. The localised field for a triplet of defects}

Following on from figure~\ref{fig:p1}, the corresponding results for a triplet of defects is presented here.
Figures~\ref{fig:p2}--\ref{fig:p2-f3} show the localised field for the case of $N=3$ with contrast ratio of $r=0.4$.
In each case, the solid curves show the displacement field, whilst the dashed curves show the associated asymptotics in the far field.
The reader is referred to section~3(\ref{subsec:triplet}) on page~\pageref{subsec:triplet} for further discussion of the figures presented here.

\begin{figure}[htb]
\centering
\subfigure[\label{fig:p2-m1}
Symmetric mode at $\omega=2.83$]{
\includegraphics[width=0.35\linewidth]{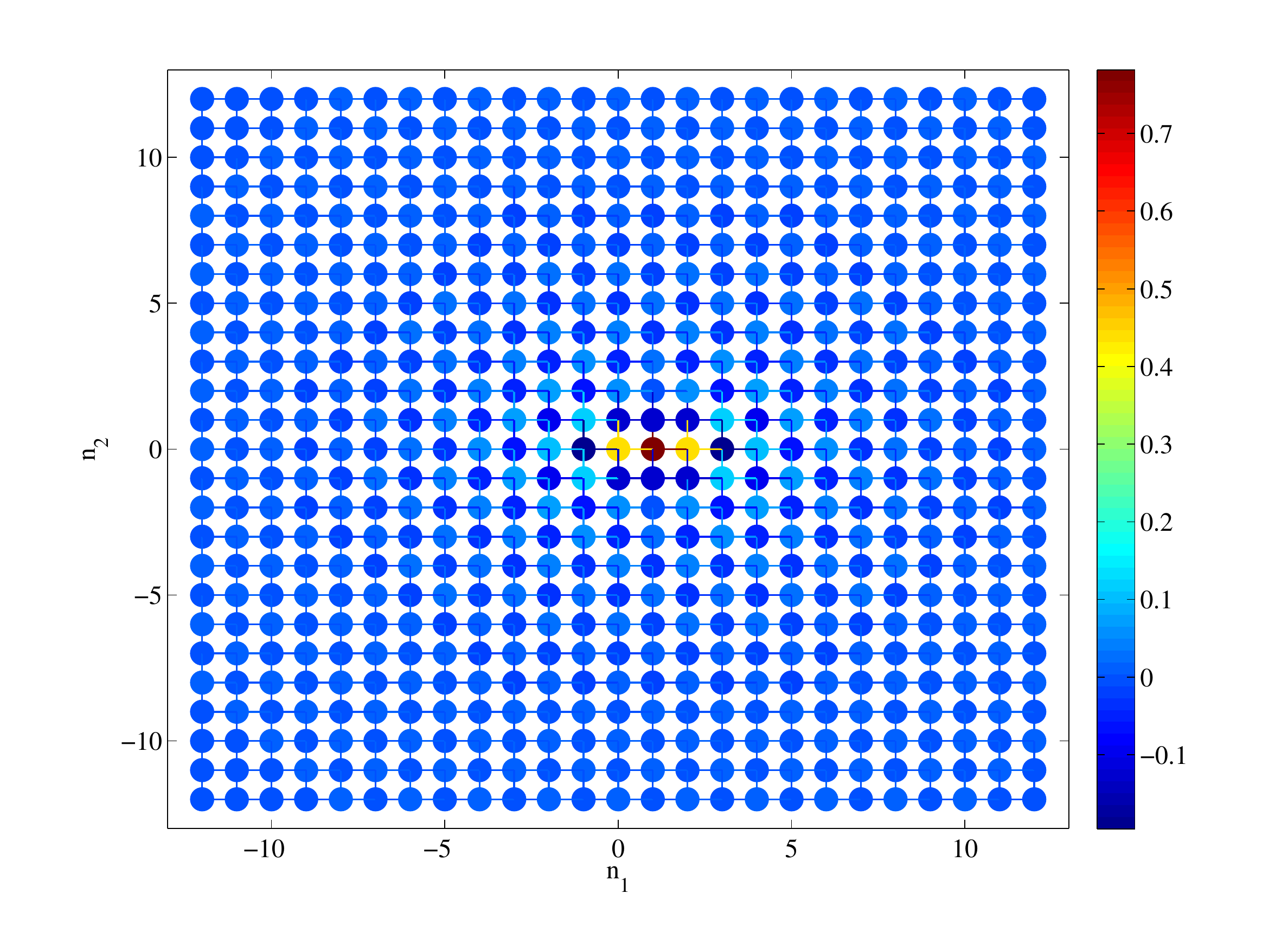}
}\qquad
\subfigure[\label{fig:p2-m1-n2-0}
The field along the line $n_2=0$ for the symmetric mode]{
\includegraphics[width=0.35\linewidth]{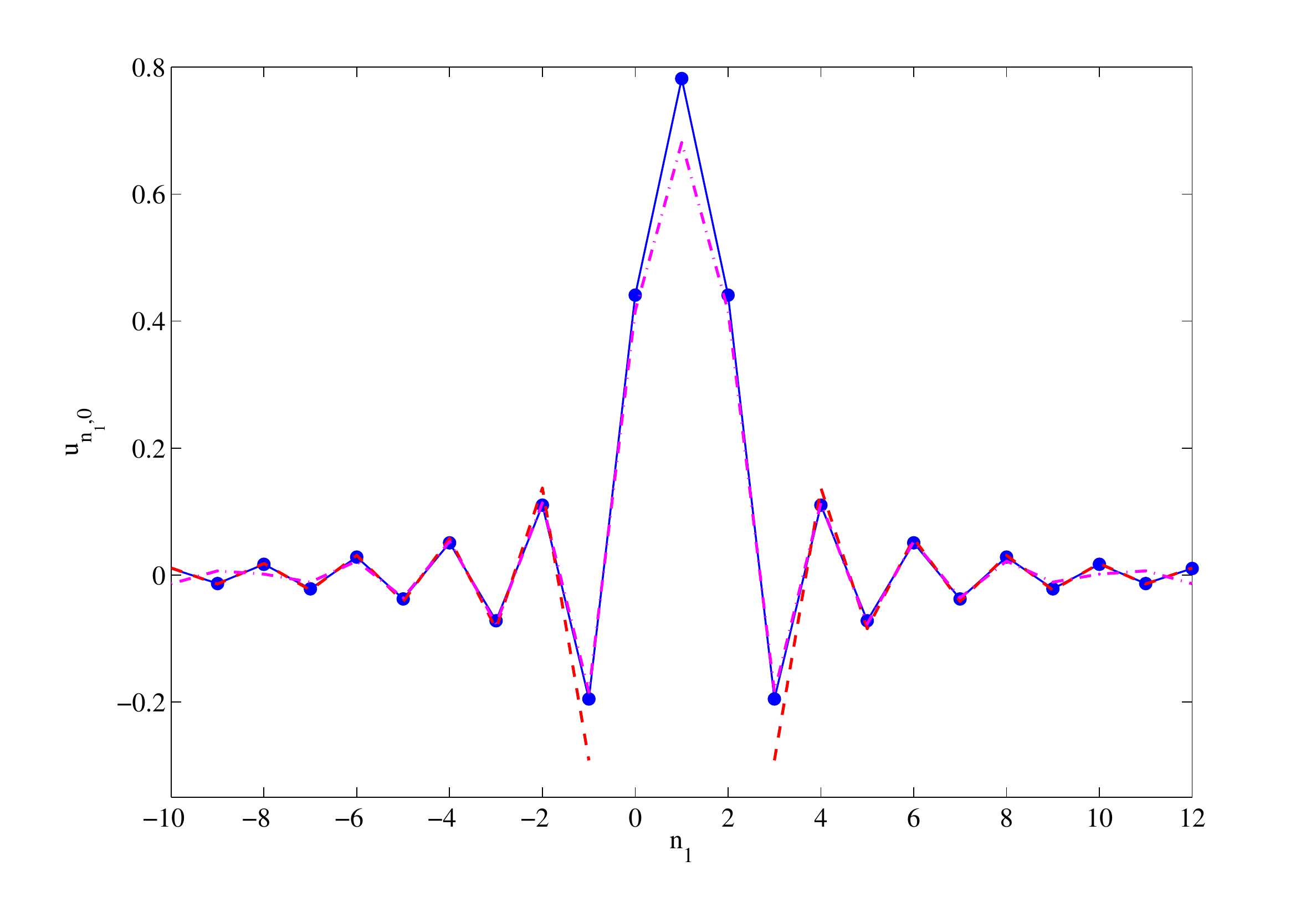}
}\qquad
\subfigure[\label{fig:p2-m1-n1-0}
The field along the line $n_1=0$ for the symmetric mode]{
\includegraphics[width=0.35\linewidth]{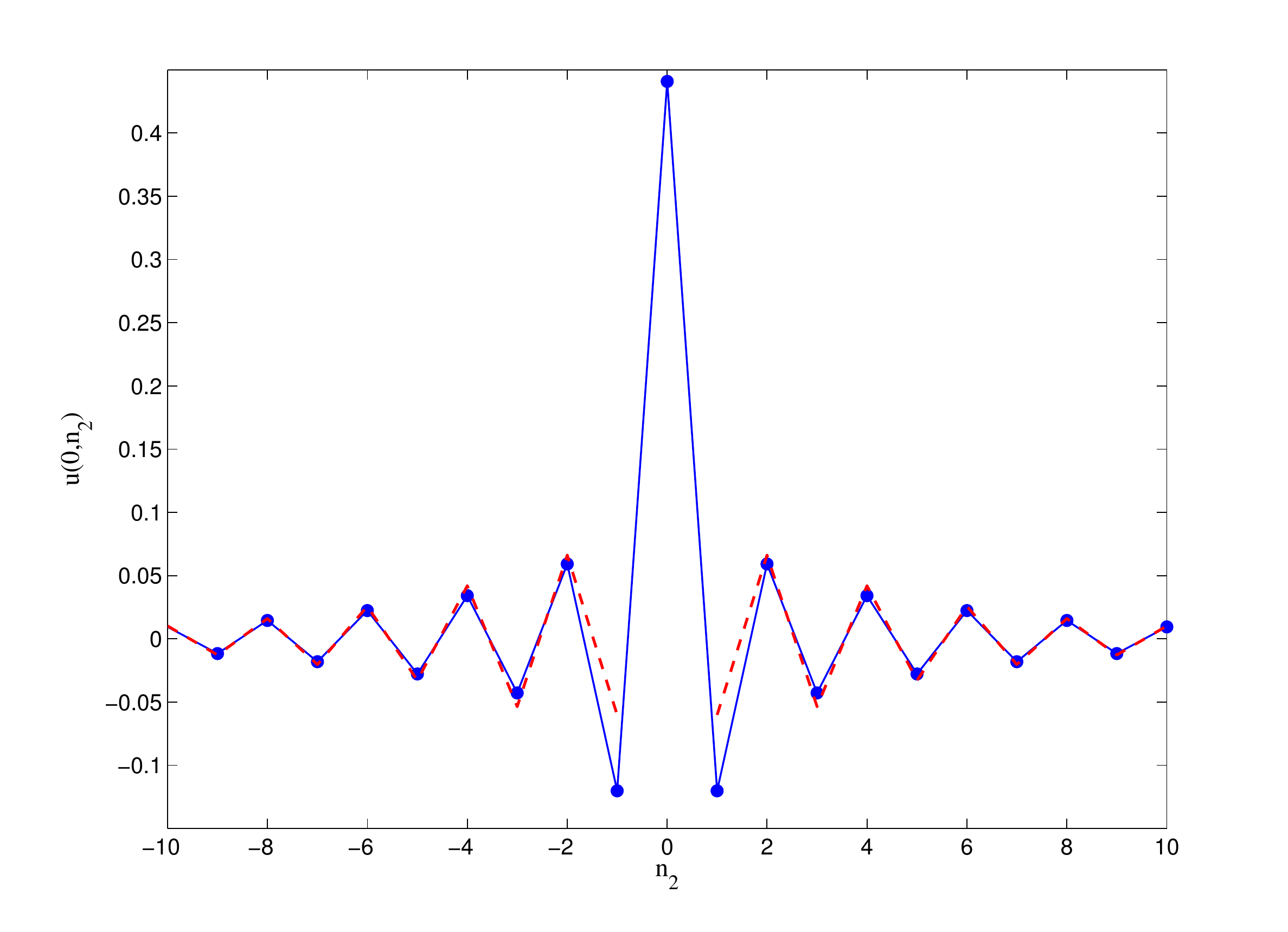}
}\qquad
\subfigure[\label{fig:p2-m1-n1-1}
The field along the line $n_1=1$ for the symmetric mode]{
\includegraphics[width=0.35\linewidth]{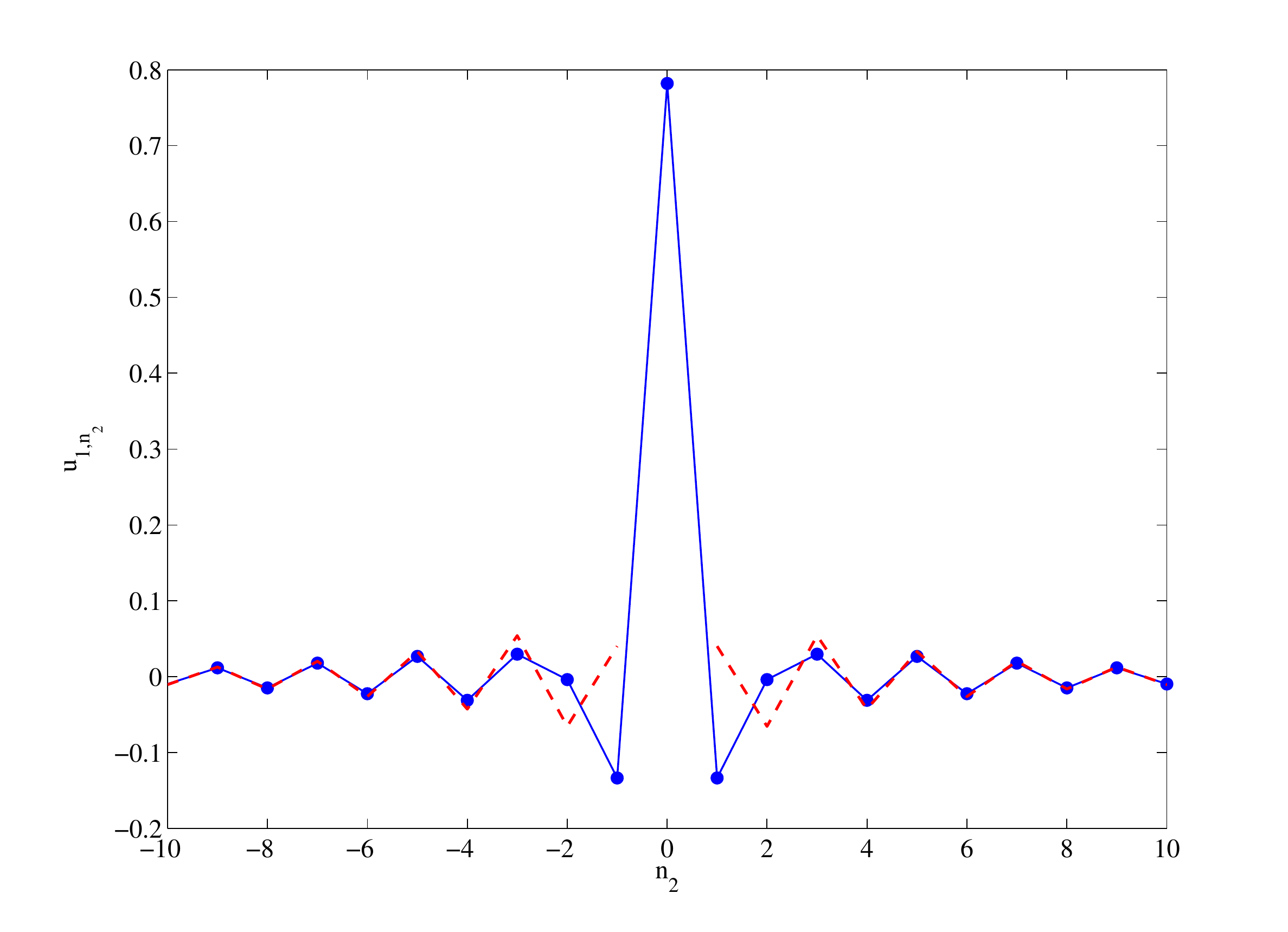}
}
\caption{\label{fig:p2}
The first localised defect mode for a triplet of defects with $r=0.4$.
The solid curves are the out-of-plane displacement along the indicated line, and the dashed curves are the associated asymptotic expansions in the far field (cf. equations~\eqref{eq:on-force-line} and~\eqref{eq:perp-force-line} as appropriate).
The dash-dot line in (b) correspond to the band edge expansion (cf. equation~\eqref{eq:band-edge-field-n2-0}).}
\end{figure}
\begin{figure}[htb]
\centering
\subfigure[\label{fig:p2-m2}
Skew-symmetric mode at $\omega=3.33$]{
\includegraphics[width=0.35\linewidth]{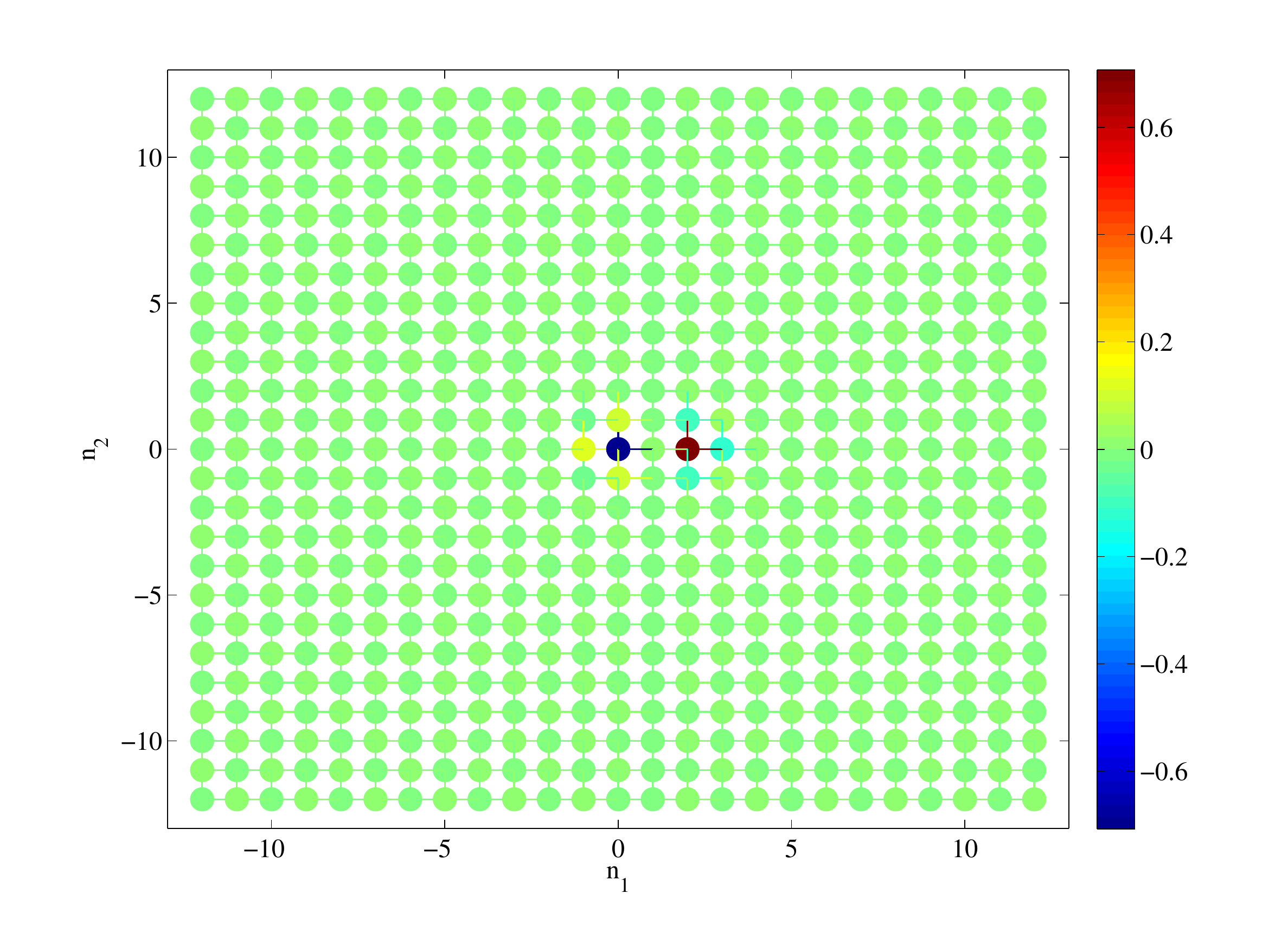}
}\qquad
\subfigure[\label{fig:p2-m2-n2-0}
The field along the line $n_2=0$ for the skew-symmetric mode]{
\includegraphics[width=0.35\linewidth]{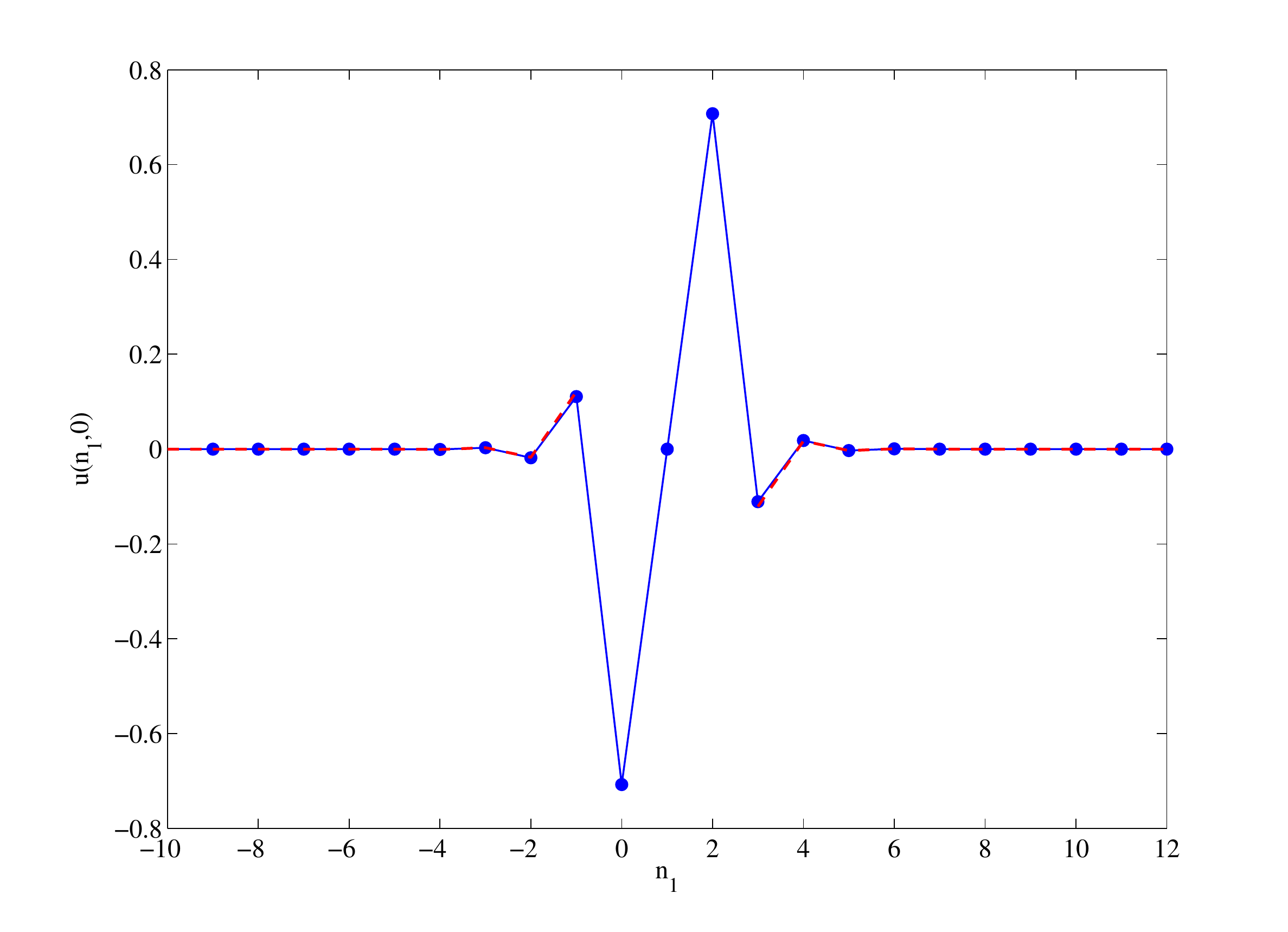}
}\qquad
\subfigure[\label{fig:p2-m2-n1-0}
The field along the line $n_1=0$ for the skew-symmetric mode]{
\includegraphics[width=0.35\linewidth]{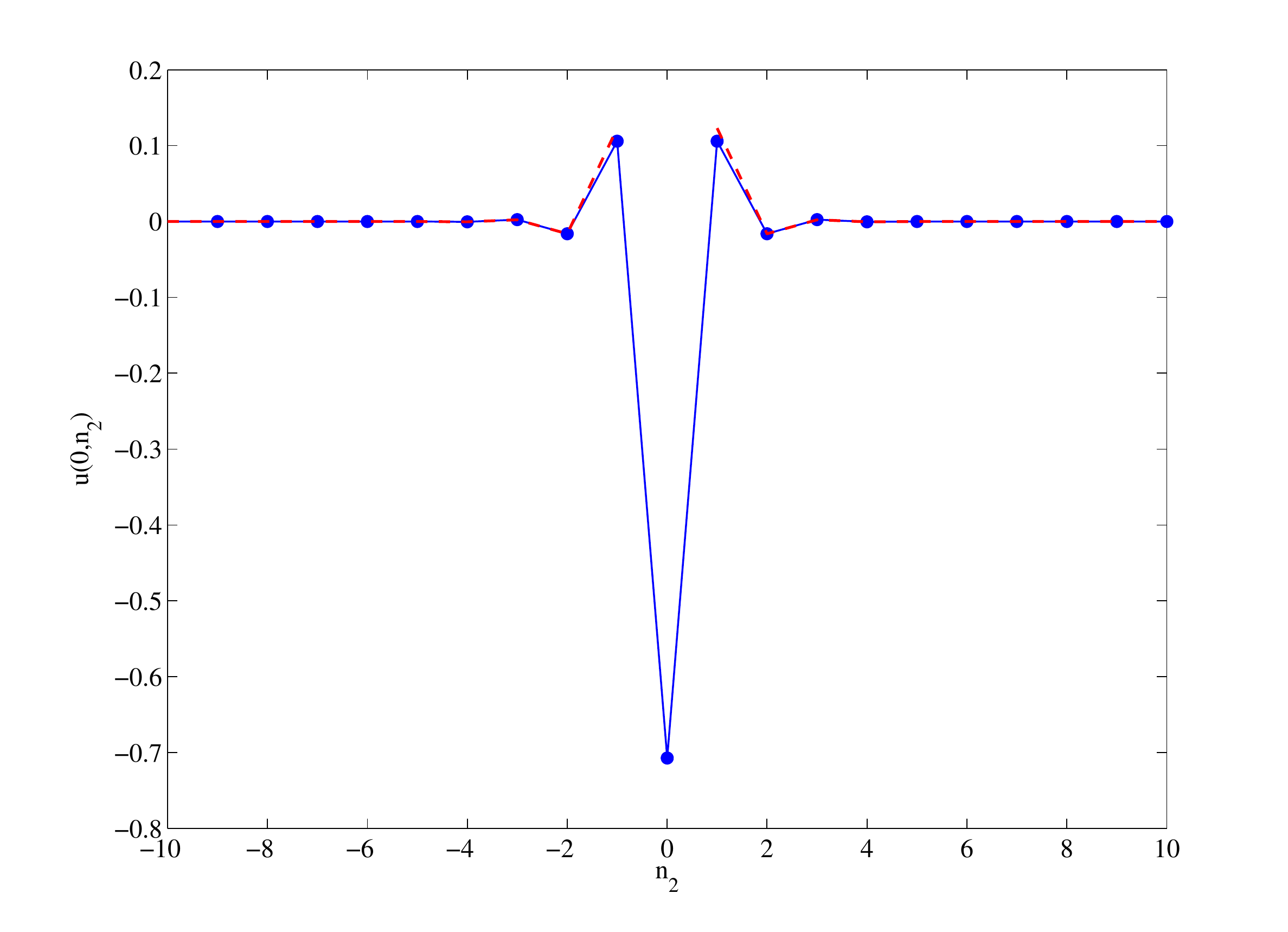}
}\qquad
\subfigure[\label{fig:p2-m2-n1-1}
The field along the line $n_1=1$ for the skew-symmetric mode]{
\includegraphics[width=0.35\linewidth]{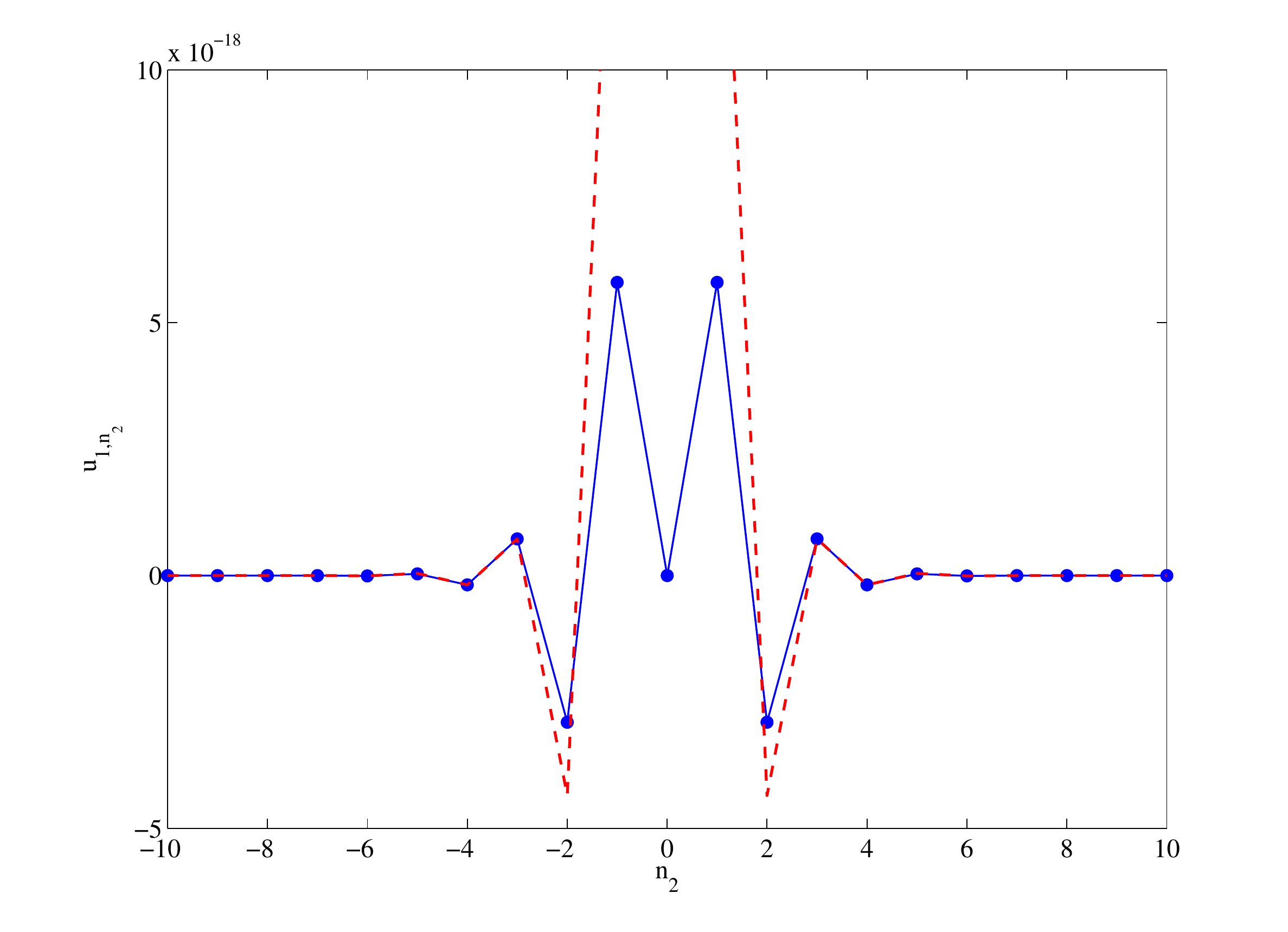}
}
\caption{\label{fig:p2-f2}
The second localised mode for a triplet of defects.
The solid curves are the out-of-plane displacement along the indicated line, and the dashed curves are the associated asymptotic expansions in the far field (cf. equations~\eqref{eq:on-force-line} and~\eqref{eq:perp-force-line} as appropriate).}
\end{figure}
\begin{figure}[htb]
\centering
\subfigure[\label{fig:p2-m3}
The second symmetric mode at $\omega=3.77$]{
\includegraphics[width=0.35\linewidth]{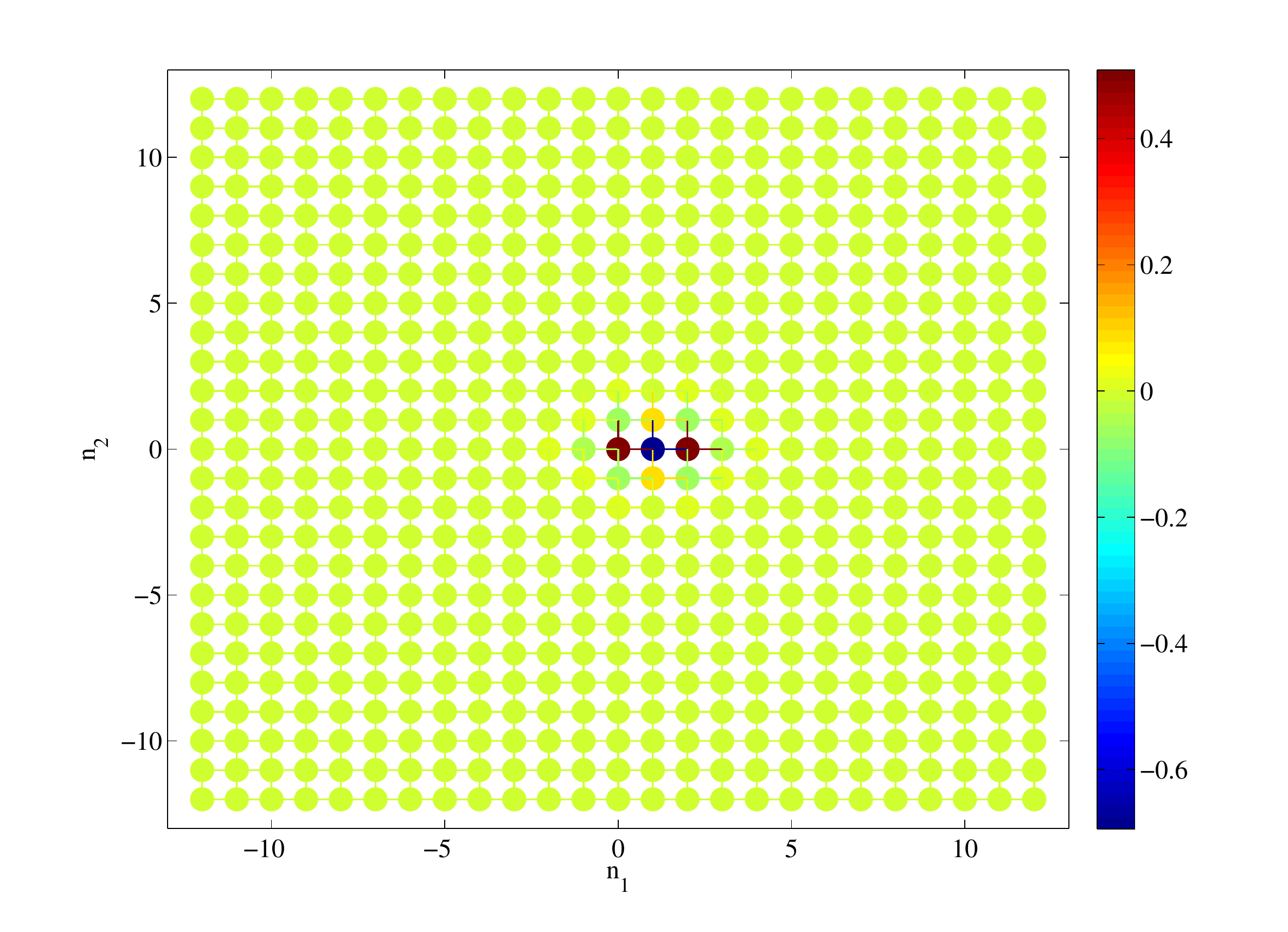}
}\qquad
\subfigure[\label{fig:p2-m3-n2-0}
The field along the line $n_2=0$ for the second symmetric mode]{
\includegraphics[width=0.35\linewidth]{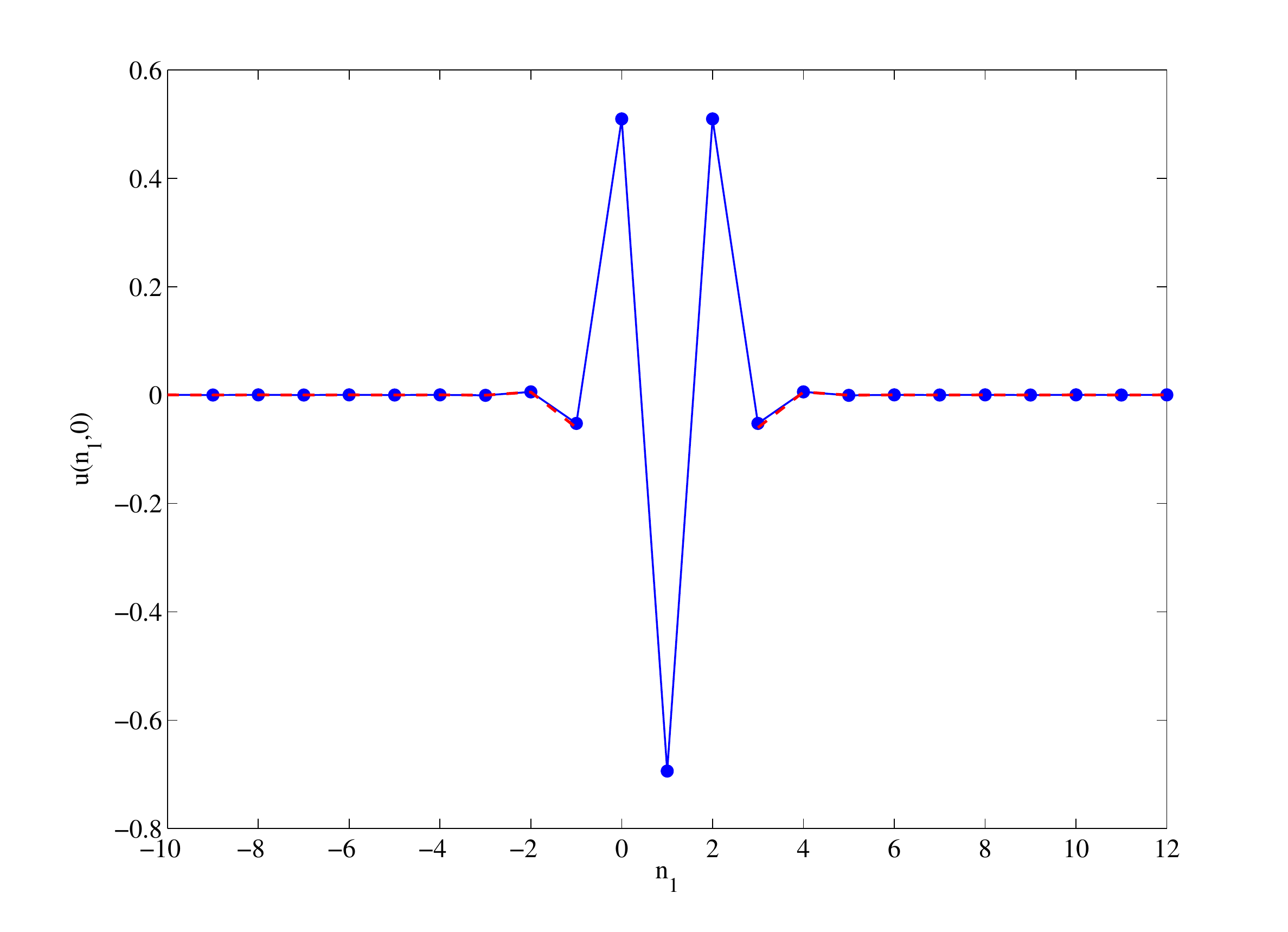}
}\qquad
\subfigure[\label{fig:p2-m3-n1-0}
The field along the line $n_1=0$ for the second symmetric mode]{
\includegraphics[width=0.35\linewidth]{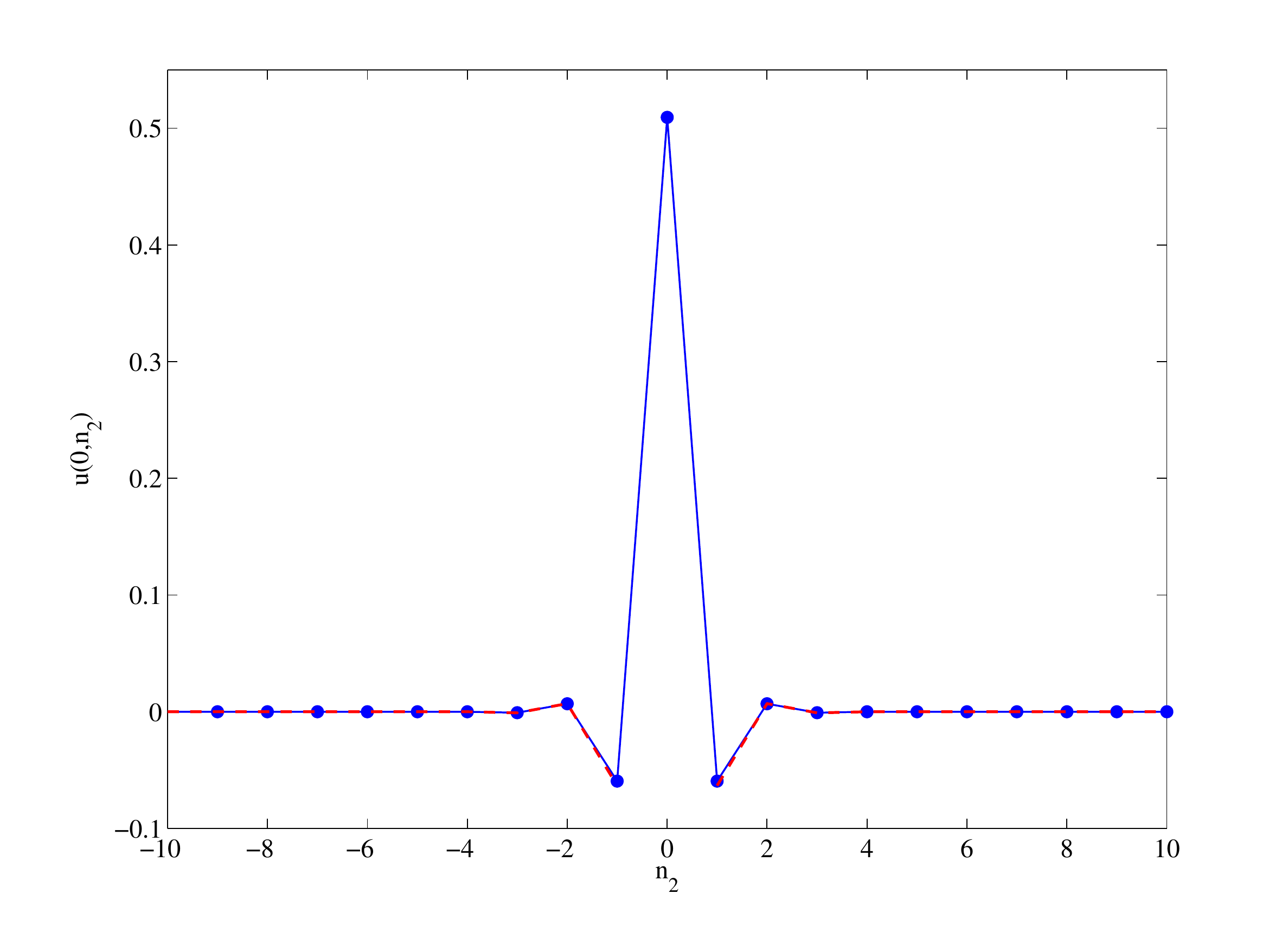}
}\qquad
\subfigure[\label{fig:p2-m3-n1-1}
The field along the line $n_1=1$ for the second symmetric mode]{
\includegraphics[width=0.35\linewidth]{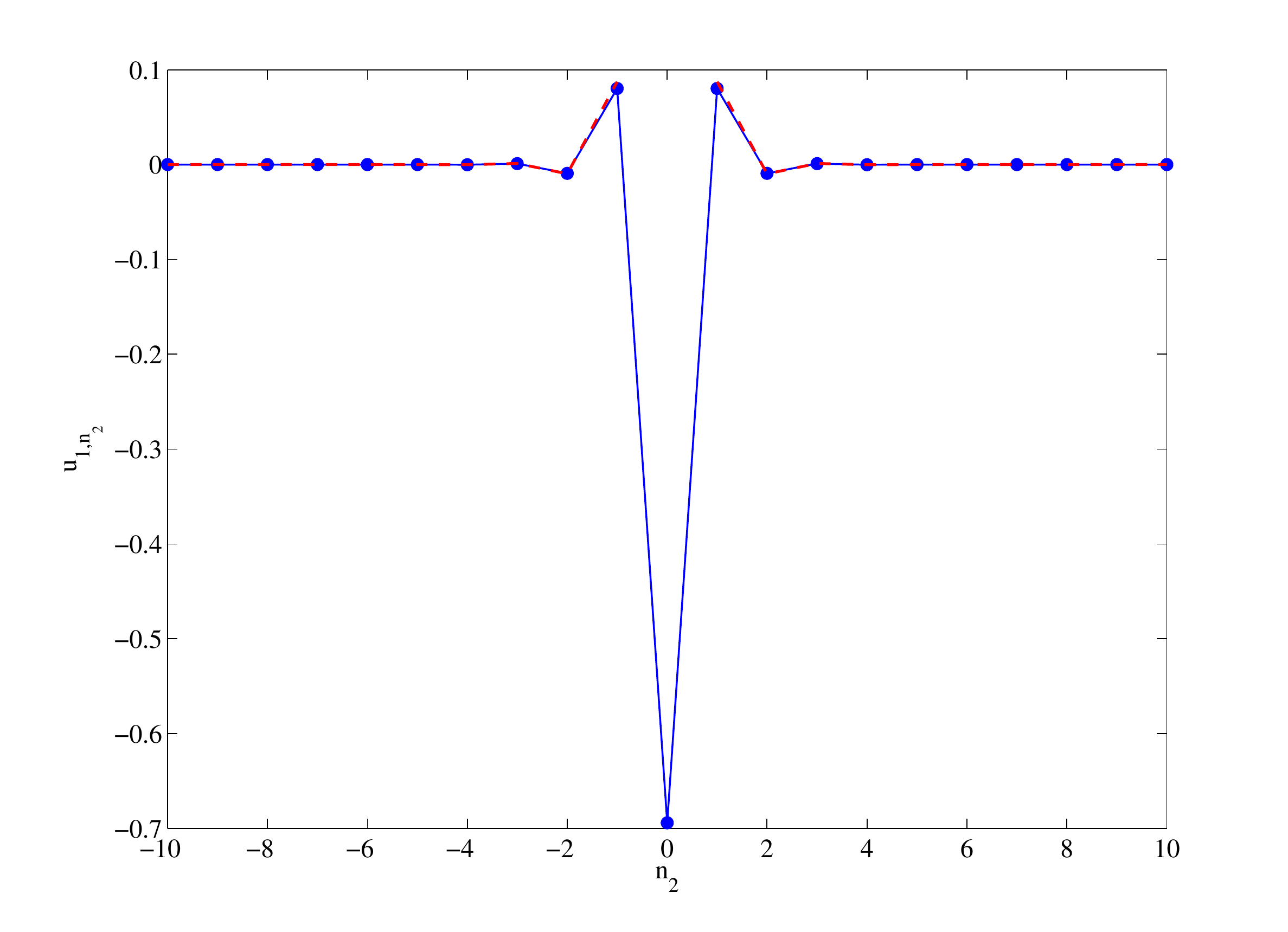}
}\qquad
\caption{\label{fig:p2-f3}
The third localised defect mode for a triplet of defects.
The solid curves are the out-of-plane displacement along the indicated line, and the dashed curves are the associated asymptotic expansions in the far field (cf. equations~\eqref{eq:on-force-line} and~\eqref{eq:perp-force-line} as appropriate).}
\end{figure}


\begin{thebibliography}{22}
\providecommand{\natexlab}[1]{#1}
\expandafter\ifx\csname urlstyle\endcsname\relax
  \providecommand{\doi}[1]{doi:\discretionary{}{}{}#1}\else
  \providecommand{\doi}{doi:\discretionary{}{}{}\begingroup
  \urlstyle{rm}\Url}\fi
\bibitem[{Ayzenberg-Stepanenko \& Slepyan(2008)}]{Ayzenbergstepanenko2008}
Ayzenberg-Stepanenko MV,  Slepyan LI. 2008 Resonant-frequency primitive
  waveforms and star waves in lattices.
\newblock \emph{J. Sound Vib.}, \textbf{313}, 812--821.
\bibitem[{B\"{u}hring(1992)}]{burhring}
B\"{u}hring W. 1992 Generalized hypergeometric functions at unit argument.
\newblock \emph{Proc. Am. Math. Soc.},
  \textbf{114},145--153.
\bibitem[{Cantoni \& Butler(1976)}]{cantoni}
Cantoni A,  Butler P. 1976 Eigenvalues and eigenvectors of symmetric
  centrosymmetric matrices.
\newblock \emph{Linear Algebr. Appl.}, \textbf{13}, 275--288.
\bibitem[{Colquitt \emph{et~al.}(2012)Colquitt, Jones, Movchan \&
  Movchan}]{Colquitt2011}
Colquitt DJ, Jones IS, Movchan NV, Movchan AB, McPhedran RC. 2012 Dynamic
  anisotropy and localization in elastic lattice systems.
\newblock \emph{Waves Random Complex Media}, \textbf{22}, 143--159.
\bibitem[{Craster \emph{et~al.}(2010)Craster, Kaplunov \&
  Postnova}]{Craster2010}
Craster RV Kaplunov J,  Postnova, J. 2010 High-frequency asymptotics,
  homogenisation and localisation for lattices.
\newblock \emph{The Q. J. Mech. Appl. Math.},
  \textbf{63}, 497--519.
\bibitem[{Delves \& Joyce(2007)}]{Delves2007}
Delves RT, Joyce, GS. 2007 Derivation of exact product forms for the
  simple cubic lattice green function using fourier generating functions and
  Lie group identities.
\newblock \emph{J. Phys. A. Math. Theor.},
  \textbf{40}, 8329--8343.
\bibitem[{Dossou \emph{et~al.}(2008)Dossou, Botten, McPhedran \&
  Poulton}]{dossou2008}
Dossou KB, Botten LC, McPhedran RC, Poulton CG. 2008 Shallow
  defect states in two-dimensional photonic crystals.
\newblock \emph{Phys. Rev. A}, \textbf{77}, 063839.
\bibitem[{Gei \emph{et~al.}(2009)Gei, Movchan \& Bigoni}]{Gei2009}
Gei M, Movchan AB,  Bigoni D. 2009 Band-gap shift and defect-induced
  annihilation in prestressed elastic structures.
\newblock \emph{J. Appl. Phys.}, \textbf{105}(6), 063507.
\bibitem[{Joyce \& Zucker(2001)}]{Joyce2001}
Joyce GS, Zucker IJ. 2001 Evaluation of the watson integral and
  associated logarithmic integral for the d-dimensional hypercubic lattice.
\newblock \emph{J. Phys. A Math. Gen}, \textbf{34},
  7349--7354.
\bibitem[{Mahmoodian \emph{et~al.}(2009)Mahmoodian, McPhedran, de~Sterke,
  Dossou, Poulton \& Botten}]{mahmoodian2009}
Mahmoodian S, McPhedran RC, de~Sterke C, Dossou KB, Poulton CG, Botten LC 2009 Single and coupled degenerate defect modes in
  two-dimensional photonic crystal band gaps.
\newblock \emph{Phys. Rev. A}, \textbf{79}, 013814.
\bibitem[{Maradudin(1965)}]{maradudin1965}
Maradudin AA. 1965 Some effects of point defects on the vibrations of
  crystal lattices.
\newblock \emph{Rep. Progr. Phys.}, \textbf{28}, 331--380.
\bibitem[{Martin(2006)}]{martin2006}
Martin PA. 2006 Discrete scattering theory: Green's function for a square
  lattice.
\newblock \emph{Wave Motion}, \textbf{43}, 619--629.
\bibitem[{Mishuris \emph{et~al.}(2009)Mishuris, Movchan \&
  Slepyan}]{Mishuris_etal}
Mishuris GS, Movchan AB, Slepyan LI. 2009 Localised knife waves in
  a structured interface.
\newblock \emph{J. Mech. Phys. Solids}, \textbf{57},
  1958--1979.
\bibitem[{Movchan \& Slepyan(2007)}]{movchan-slepyan}
Movchan AB, Slepyan LI. 2007 Band gap green's functions and localized
  oscillations.
\newblock \emph{Proc. R. Soc. A}, \textbf{463}, 2709--2727.
\bibitem[{Newton(1687)}]{Newton}
Newton I. 1687 \emph{Principia}.
\newblock London, UK: The Royal Society.
\bibitem[{Nieves \emph{et~al.}(2012)Nieves, Movchan, Jones \&
  Mishuris}]{Nieves2012}
Nieves MJ, Movchan AB, Jones IS, Mishuris GS. 2012
  {Propagation of Slepyan's crack in a non-uniform elastic lattice}.
\newblock (\url{http://arxiv.org/abs/1204.57766}).
\bibitem[{Osharovich \& Ayzenberg-Stepanenko(2012)}]{Osharovich_etal}
Osharovich GG, Ayzenberg-Stepanenko MV. 2012 Wave localization in
  stratified square-cell lattices: The antiplane problem.
\newblock \emph{J. Sound Vib.}, \textbf{331}, 1378--1397.
\bibitem[{Prudnikov \emph{et~al.}(1992)Prudnikov, Brychkov \&
  Marichev}]{prudnikov-v4}
Prudnikov AP, Brychkov YA, Marichev OI 1992 \emph{Integrals and
  series}, vol.~4.
\newblock Amsterdam, The Netherlands: Gordon and Breach Science Publishers.
\bibitem[{Saigo \& Srivastava(1990)}]{saigo}
Saigo M,  Srivastava HM. 1990 The behavior of the zero-balanced
  hypergeometric series {${}_{p}\mathrm{F}_{p-1}$} near the boundary of its
  convergence region.
\newblock \emph{Proc. A. Math. Soc.},
  \textbf{110}, 71--76.
\bibitem[{Slepyan(2002)}]{Slepyan2002}
Slepyan LI. 2002 \emph{Models and phenomena in fracture mechanics}.
\newblock Berlin, Germany: Springer.
\bibitem[{{van der Pol} \& Bremmer(1950)}]{van-der-pol}
{van der Pol}, B. \& Bremmer, H. 1950 \emph{{Operational Calculus based on the
  two-sided Laplace Transform}}.
\newblock London: Cambridge University Press.
\bibitem[{Zucker(2011)}]{Zucker2011}
Zucker IJ. 2011 {70+ Years of the Watson Integrals}.
\newblock \emph{Journal of Statistical Physics}, \textbf{145}, 591--612.

\end{thebibliography}
\end{document}